\documentclass[aps,prd,reprint,superscriptaddress,floatfix]{revtex4-1}

\usepackage[utf8x]{inputenc}
\usepackage{microtype}

\usepackage{graphicx}
\usepackage[dvipsnames]{xcolor}
\usepackage{rotating}

\usepackage{amsmath}
\usepackage{amssymb}
\usepackage{amsfonts,dsfont}
\usepackage{bm}

\usepackage{tabu}
\usepackage{booktabs}
\usepackage{arydshln}
\usepackage{multirow}
\usepackage{array}
\usepackage{enumitem}
\usepackage{tensor}

\usepackage{url}
\usepackage[colorlinks=true,urlcolor=blue,linkcolor=blue,citecolor=blue]{hyperref}

\newcommand {\apgt} {\ {\raise-.5ex\hbox{$\buildrel>\over\sim$}}\ }
\newcommand {\aplt} {\ {\raise-.5ex\hbox{$\buildrel<\over\sim$}}\ }

\def\s0#1#2{\mbox{\small{$ \frac{#1}{#2} $}}}
\def\0#1#2{\frac{#1}{#2}}

\def\CC{{\mathcal C}}

\def\CN{{\mathcal N}}

\definecolor{darkgreen}{rgb}{0,0.6,0}
\definecolor{gray}{rgb}{.7,.7,.7}

\DeclareMathOperator{\tr}{Tr}
\DeclareSymbolFont{symbolsC}{U}{pxsyc}{m}{n}
\DeclareMathSymbol{\coloneqq}{\mathrel}{symbolsC}{"42}
\DeclareMathAlphabet{\EuRoman}{U}{eur}{m}{n}
\SetMathAlphabet{\EuRoman}{bold}{U}{eur}{b}{n}

 \graphicspath{{./plots/}}

\begin{document}

\newcommand{\background}[1]{\bar{#1}}
\newcommand{\backgroundg}{\background{g}}
\newcommand{\backgroundLambda}{\background{\lambda}}
\newcommand{\imaginaryi}{\mathrm{i}}

\newcommand{\eulere}{\mathrm{e}}
\newcommand{\vect}[1]{\bm{#1}}
\newcommand{\tens}[1]{#1}
\newcommand{\scaleDerivative}[1]{\partial_t #1}
\renewcommand{\d}[1]{\mathop{\mathrm{d}#1}}
\newcommand{\dn}[2]{\mathop{\mathrm{d}^{#1} #2}}
\newcommand{\D}[1]{\mathcal{D}#1}
\newcommand{\abs}[1]{\left|#1\right|}
\newcommand{\proj}{\Pi}
\newcommand{\functionalDerivative}[2]{\frac{\delta #2}{\delta #1}}
\newcommand{\contract}{\circ}
\newcommand{\expectationValue}[1]{\langle #1 \rangle}

\newcommand{\definition}{\coloneqq}
\newcommand{\identity}{\delta}
\newcommand{\manifold}{\mathcal{M}}
\newcommand{\metric}{{g}}
\newcommand{\metricBackground}{\background{\metric}}
\newcommand{\graviton}{h}
\newcommand{\sqrtg}{\sqrt{\metric}}
\newcommand{\sqrtgbg}{\sqrt{\metricBackground}}
\newcommand{\ThetaFunction}{\Theta}
\newcommand{\ttSymb}{\text{\tiny TT}}
\newcommand{\ltSymb}{\text{\tiny LT}}
\newcommand{\llSymb}{\text{\tiny LL}}
\newcommand{\tSymb}{\text{\tiny T}}
\newcommand{\projGn}[1]{\proj_{G_n}}
\newcommand{\projLambdan}[1]{\proj_{\cosmologicalConstant_n}}
\newcommand{\projTT}{\proj_\ttSymb}

\newcommand{\gravitonTT}{\graviton^{\ttSymb}}
\newcommand{\gravitonLT}{\graviton^{\ltSymb}}
\newcommand{\gravitonLL}{\graviton^{\llSymb}}
\newcommand{\gravitonTr}{\graviton^{\tr}}
\newcommand{\Mass}{M}
\newcommand{\Masssquared}{\Mass^2}
\newcommand{\mass}{\mu}
\newcommand{\massPlanck}{M_{\text{Pl}}}

\DeclareRobustCommand{\order}{\ensuremath{\mathcal{O}}}

\newlength{\leftstackrelawd}
\newlength{\leftstackrelbwd}
\def\leftstackrel#1#2{\settowidth{\leftstackrelawd}%
{${{}^{#1}}$}\settowidth{\leftstackrelbwd}{$#2$}%
\addtolength{\leftstackrelawd}{-\leftstackrelbwd}%
\leavevmode\ifthenelse{\lengthtest{\leftstackrelawd>0pt}}%
{\kern-.5\leftstackrelawd}{}\mathrel{\mathop{#2}\limits^{#1}}}

\newcommand{\refequal}[1]{\stackrel{\labelcref{#1}}{=}}
\newcommand{\refequalaligned}[1]{\leftstackrel{\labelcref{#1}}{=}}

\newcommand{\wfr}[3][]{\wfrsymb^{#1}_{#2} (#3)}
\newcommand{\wfrsymb}{Z}
\newcommand{\anomalousDimensionSymb}{\eta}
\newcommand{\anomalousDimension}[1]{\anomalousDimensionSymb_{#1}}
\newcommand{\anomalousDimensionConst}[1]{\hat{\anomalousDimensionSymb}_{#1}}
\newcommand{\R}{R}
\newcommand{\RicciTensor}{\tens{R}}
\newcommand{\RiemannTensor}[1]{\tensor{\tens{R}}{#1}}
\newcommand{\ChristoffelSymbol}[1]{\tensor*{\Gamma}{#1}}
\newcommand{\metricTensor}[1]{\tensor{\metric}{#1}}
\newcommand{\covariantDerivative}{\nabla}
\newcommand{\cosmologicalConstant}{\Lambda}
\newcommand{\ghost}{c}
\newcommand{\antighost}{\bar{\ghost}}
\newcommand{\cutoff}{\mu}
\newcommand{\DiracConj}[1]{\bar{#1}}
\newcommand{\nptfct}[1]{$#1$-point function}
\newcommand{\regulator}{R}
\newcommand{\regulatorOfField}[1]{\regulator^{#1}}
\newcommand{\dROfField}[1]{\scaleDerivative{\regulator}^{#1}}
\newcommand{\shapeFunction}{r}
\newcommand{\RGtime}{t}
\newcommand{\energyMomentumTensor}{T}

\newcommand{\flow}{\text{Flow}}

\newcommand{\fofR}{f(\R)}

\newcommand{\lagrangian}{\mathcal{L}}
\newcommand{\lagrangianMatter}{\lagrangian_\text{matter}}
\newcommand{\actionCl}{S}
\newcommand{\actionQu}{\Gamma}
\newcommand{\actionEff}[1]{\Gamma_{#1}}
\newcommand{\actionReg}[1]{\Delta\actionCl_{#1}}
\newcommand{\actionEH}{\actionCl_\text{EH}}
\newcommand{\actionMatter}{\actionCl_\text{matter}}
\newcommand{\actionfofR}{\actionCl_{\fofR}}
\newcommand{\actionGF}{\actionCl_\text{gf}}
\newcommand{\actionGhost}{\actionCl_\text{gh}}
\newcommand{\tensorStructure}{\tens{\mathcal{T}}}
\newcommand{\FPoperator}{\mathcal{M}}
\newcommand{\partitionFunctional}{\mathcal{Z}}
\newcommand{\WignerFunctional}{\mathcal{W}}
\newcommand{\vertex}[1]{\Gamma^{(#1)}}

\newcommand{\functional}{f}

\newcommand{\correlatorDisc}[2]{\langle #2 \rangle_{#1}}
\newcommand{\propagator}{\mathcal{G}}
\newcommand{\propagatorOfField}[1]{\propagator^{#1}}
\newcommand{\source}{J}
\newcommand{\superfield}[1]{#1}
\newcommand{\superfieldFluctuating}{\superfield{\phi}}

\newcommand{\newtonG}{G_N}
\newcommand{\Gn}[1]{G_{#1}}
\newcommand{\gn}[1]{g_{#1}}
\newcommand{\gnd}[1]{\scaleDerivative{g}_{#1}}
\newcommand{\Lambdan}[1]{\cosmologicalConstant_{#1}}
\newcommand{\lambdan}[1]{\lambda_{#1}}
\newcommand{\lambdand}[1]{\scaleDerivative{\lambda}_{#1}}
\newcommand{\constantSymb}{C}
\newcommand{\constantLambdan}[1]{\constantSymb^{\Lambdan{#1}}}
\newcommand{\constantGnp}[1]{\constantSymb^{\Gn{#1}}_{p^2}}
\newcommand{\constantGnLambda}[1]{\constantSymb^{\Gn{#1}}_{\Lambdan{#1}}}
\newcommand{\constantGnRsquared}[1]{\constantSymb^{\Gn{#1}}_{\RsquaredCouplingn{#1}}}
\newcommand{\RsquaredCouplingdim}{\Omega}
\newcommand{\RsquaredCouplingndim}[1]{\RsquaredCouplingdim_{#1}}
\newcommand{\RsquaredCoupling}{\omega}
\newcommand{\RsquaredCouplingn}[1]{\RsquaredCoupling_{#1}}

\newcommand{\coupling}{\alpha}
\newcommand{\couplingg}[1]{\coupling_{#1}}
\newcommand{\couplinggd}[1]{\scaleDerivative{\coupling}_{#1}}
\newcommand{\couplinggPert}[1]{\Delta\couplingg{#1}}
\newcommand{\fpstar}[1]{{#1}^\ast}
\newcommand{\stabilityMatrix}{B}
\newcommand{\stabilityMatrixAlt}{\widetilde \stabilityMatrix}
\newcommand{\criticalExponent}{\theta}
\newcommand{\eigenvector}{e}

\newcommand{\betafct}{\beta}
\newcommand{\renormalisationScale}{\cutoff}

\newcommand{\gaugeFixingCondition}{\mathcal{F}}
\newcommand{\gfone}{\xi_1}
\newcommand{\gftwo}{\xi_2}

\newcommand{\symmetryGroup}{\mathcal{G}}

\title{Towards apparent convergence in asymptotically safe quantum gravity}

\author{T.~Denz}
\affiliation{Institut für Theoretische Physik, Universit\"at Heidelberg,
Philosophenweg 16, 69120 Heidelberg, Germany}
\author{J.~M.~Pawlowski}
\affiliation{Institut für Theoretische Physik, Universit\"at Heidelberg,
Philosophenweg 16, 69120 Heidelberg, Germany}
\affiliation{ExtreMe Matter Institute EMMI, GSI Helmholtzzentrum für
Schwerionenforschung mbH, Planckstr.\ 1, 64291 Darmstadt, Germany}
\author{M.~Reichert}
\affiliation{Institut für Theoretische Physik, Universit\"at Heidelberg,
Philosophenweg 16, 69120 Heidelberg, Germany}

\begin{abstract}
The asymptotic safety scenario in gravity is accessed within the
systematic vertex expansion scheme for functional renormalisation
group flows put forward in
\cite{Christiansen:2012rx,Christiansen:2014raa}, and implemented in
\cite{Christiansen:2015rva} for propagators and three-point
functions. In the present work this expansion scheme is extended to
the dynamical graviton four-point function. For the first time, this
provides us with a closed flow equation for the graviton propagator:
all vertices and propagators involved are computed from their own
flows. 

In terms of a covariant operator expansion the current approximation
gives access to $\Lambda$, $R$, $R^2$ as well as $R_{\mu\nu}^2$ and
higher derivative operators. We find a UV fixed point with three
attractive and two repulsive directions, thus confirming previous
studies on the relevance of the first three operators. In the
infrared we find trajectories that correspond to classical general
relativity and further show non-classical behaviour in some
fluctuation couplings. 

We also find signatures for the
apparent convergence of the systematic vertex expansion. This opens
a promising path towards establishing asymptotically safe
gravity in terms of apparent convergence.

\end{abstract}

\maketitle

\section{Introduction}
\label{sec:intro}
A consistent formulation of quantum gravity (QG) is an open problem,
with many contenders being investigated in great detail. In the past
two decades, Weinberg's asymptotic safety scenario (AS) proposed in
1976 \cite{Weinberg:1980gg} has been investigated with the help of
non-perturbative renormalisation group techniques.  AS posits that QG,
while being perturbatively non-renormalisable, is non-perturbatively
renormalisable and features a non-trivial fixed point in the
ultraviolet (UV).  

By now an impressive body of evidence has been collected that supports
this intriguing scenario: the application of functional
renormalisation group (FRG) techniques \cite{Wetterich:1992yh} to 
QG \cite{Reuter:1996cp} has allowed for a confirmation of the existence
of the non-trivial UV fixed point in basic Einstein-Hilbert
approximations \cite{Reuter:1996cp,Souma:1999at,Reuter:2001ag}.  Later
works have improved on these approximations and gone beyond
Einstein-Hilbert
\cite{Lauscher:2001ya,Lauscher:2002sq,Litim:2003vp,Fischer:2006fz,Codello:2006in,
  Codello:2007bd,Machado:2007ea,Codello:2008vh,
  Niedermaier:2009zz,Benedetti:2009rx,Eichhorn:2009ah,Manrique:2009uh,Narain:2009fy,
  Eichhorn:2010tb,Groh:2010ta,Manrique:2010am,Manrique:2011jc,
  Donkin:2012ud,Nagy:2012rn,Benedetti:2012dx,Rechenberger:2012pm,Christiansen:2012rx,Dietz:2012ic,
  Falls:2013bv,Benedetti:2013jk,Codello:2013fpa,Dietz:2013sba,Nagy:2013hka,Ohta:2013uca,
  Christiansen:2014raa,Becker:2014qya,Falls:2014zba,Falls:2014tra,
  Falls:2015qga,Eichhorn:2015bna,Falls:2015cta,Christiansen:2015rva,
  Ohta:2015efa,Gies:2015tca,Benedetti:2015zsw,Ohta:2015fcu,
  Gies:2016con,Falls:2016wsa,Falls:2016msz,Biemans:2016rvp,Pagani:2016dof,Christiansen:2016sjn}; 
for an overview see reviews
\cite{Niedermaier:2006wt,Percacci:2007sz,Litim:2011cp,Reuter:2012id}.
Furthermore, the stability of the asymptotic safety setting for
gravity-matter systems, \cite{Dou:1997fg,Percacci:2002ie,Daum:2010bc,
  Folkerts:2011jz,Harst:2011zx,Eichhorn:2011pc,
  Eichhorn:2012va,Dona:2012am,Henz:2013oxa,Dona:2013qba,
  Percacci:2015wwa,Labus:2015ska,Oda:2015sma,Meibohm:2015twa,Dona:2015tnf,
  Meibohm:2016mkp,Eichhorn:2016esv,Henz:2016aoh,Eichhorn:2016vvy}, as
well as asymptotically safe theories within a perturbative setup,
\cite{Litim:2014uca,Esbensen:2015cjw,Codello:2016muj,Bond:2016dvk,Molgaard:2016bqf,Bajc:2016efj},
have also attracted a lot of attention.

In \cite{Christiansen:2015rva} the systematic vertex expansion in
quantum gravity initiated in
\cite{Christiansen:2012rx,Christiansen:2014raa}, see also
\cite{Donkin:2012ud,Codello:2013fpa}, has been pushed to the graviton three-point
function, for the first time including a dynamical graviton-scattering
in the asymptotic safety analysis. Here we extend the vertex expansion
to the graviton four-point function. Apart from the significant
technical challenge such an upgrade of the approximation has posed, we
think that this constitutes a necessary and significant progress
towards asymptotically safe gravity:

\begin{itemize} 
\item As such it is an important step towards apparent convergence of
  the vertex expansion in quantum gravity: apparent convergence aims
  at the convergence of vertices as well as observables in the order
  of a given systematic expansion scheme; here we use the vertex
  expansion scheme. Together with the investigation of the regulator
  (in-)dependence of observables this provides a systematic error
  estimate in the present approach, and should be compared with
  apparent continuum scaling and extrapolation on the lattice.

\item The present approximation allows for the identification of
  diffeomorphism-invariant structures in the vertex expansion, i.e.\
  $\R^2$ and  $\R_{\mu\nu}^2$ tensor structures as well as those of higher derivative invariants.
  This is not only important for getting access to the
  number of relevant directions at the asymptotically safe UV fixed
  point in quantum gravity, but can also provide non-trivial support
  and additional information for computations within the standard
  background field approximation.

\item It is the first approximation in the asymptotic safety approach
  to gravity where the flow of the pivotal building block,
  the two-point correlation function or (inverse) propagator, 
  is closed: The flows of all involved vertex functions are
  computed within given approximations.  As the propagator is the core
  object in the present approach, we consider this an important
  milestone on the way towards asymptotically safe quantum gravity.

\end{itemize}

As a main result of this paper, we find further significant evidence
for a non-trivial UV fixed point in quantum gravity. This fixed point
has three relevant directions and two repulsive ones. The three
relevant directions can be associated with the cosmological constant
(graviton mass parameter), Newton's coupling and the $\R^2$-coupling,
see \autoref{sec:AS}. We also investigate the stability of this UV
fixed point and observe that the system is significantly less
sensitive to the closure of the flow equations than previous
truncations. In addition, we observe that the critical exponents
  also become less sensitive to the details of the approximations.
  These are two necessary signatures of apparent
convergence. Furthermore, we investigate the infrared (IR) behaviour
of the system and find trajectories connecting the UV fixed point with
classical general relativity in the IR.

This work is structured in the following way: In \autoref{sec:Diff} we
elaborate on the important property of background independence of
quantum gravity and its consequences for the dynamical correlation
functions. In \autoref{sec:frg}, we briefly introduce the functional
renormalisation group and the covariant vertex expansion used in the
present work. Next, \autoref{sec:flows} presents our setup and the
derivations of the flow equations for the couplings.  We especially
focus on how the tensor structures of higher curvature invariants are
embedded in our vertex expansion.  In \autoref{sec:AS}, we present our
results, in particular the non-trivial UV fixed point with three
attractive and two repulsive directions.  Furthermore, we test the
stability of the UV fixed point with respect to change of the precise
truncation, and find that in almost all approximations we obtain a
similar UV fixed point.  In \autoref{sec:IR-behaviour}, we investigate
the IR-behaviour of the UV-finite trajectories. Our analysis confirms
the classical IR fixed points found before in the vertex expansion. In
\autoref{sec:convergence} we discuss the convergence of the present
vertex expansion scheme with an increasing number of flowing $n$-point
correlation functions.  Finally, in \autoref{sec:summary} we summarise
our results.

\section{Diffeomorphism invariance and background independence}
\label{sec:Diff}
The quantum field theoretical formulation of quantum gravity in terms
of metric correlation functions necessitates the introduction of a
background metric $\bar g$, and quantum fluctuations are taken to be
fluctuations about this background metric. This begs the question of
whether diffeomorphism invariance and background independence of
observables are guaranteed in such a framework. While this is an
important question, its answer is not directly relevant for the
computations presented here. Hence, the present chapter may be skipped
in a first reading.

In the present work we perform computations in the linear split, where
the full metric $g$ is given by $g=\bar g+h$. More general splits,
$g=\bar g+f(\bar g,h)$ have been considered for example within the
geometrical or Vilkovisky-deWitt approach,
e.g.~\cite{Branchina:2003ek,Pawlowski:2003sk,Donkin:2012ud,Demmel:2014hla},
or the exponential split,
e.g.~\cite{Kawai:1992np,Nink:2014yya,Percacci:2015wwa,Falls:2015qga,Demmel:2015zfa,Gies:2015tca}.
Observables, on the other hand, are background independent. This
property is encoded in the Nielsen (NI) or split-Ward identities (SWI)
that relate derivatives of the effective action $\Gamma[\bar g,\phi]$
w.r.t.\ the background metric $\bar g$ to those w.r.t.\ the graviton fluctuations
$h$. Here we have introduced the fluctuation superfield 
\begin{align}\label{eq:phi}
\phi=(h,c,\bar c)\,. 
\end{align} 
The (anti-) ghost fields, $c$ and $\bar c$, stem from the Faddeev-Popov gauge fixing
procedure, see next section. The effective action generates all
one-particle-irreducible correlation functions and as such encodes the
symmetries of the theory. Schematically, these identities read
\cite{Reuter:1996cp,Pawlowski:2003sk,Pawlowski:2005xe,Manrique:2009uh,Manrique:2010mq,
  Donkin:2012ud,Bridle:2013sra,Dietz:2015owa,Safari:2015dva,Morris:2016spn,Percacci:2016arh}
\begin{align}\label{eq:NI}
  \0{\delta \Gamma[\bar g,h]}{\delta \bar g(x)} = \int_{y}\CC[\bar g,
  h](x,y)\0{\delta \Gamma[\bar g, h]}{\delta h(y)}+\CN[\bar g,
  h](x)\,, 
\end{align}
where we have suppressed the ghost fields. In the linear
split we have $\CC[\bar g, h](x,y)=\delta(x-y)$ and the second term
$\CN[\bar g, h]$ carries the information about the non-trivial
behaviour under diffeomorphism transformations of the gauge fixing
sector and the regularisation. In turn, in the geometrical approach
diffeomorphism invariance of the effective action is achieved by a
non-linear split with $f(\bar g,h)$ leading to the non-trivial
prefactor $\CC[\bar g, h]$ in \eqref{eq:NI}. The term $\CN[\bar g, h]$
then carries the deformation of the Nielsen identity in the presence
of a regularisation but does not spoil diffeomorphism invariance.

In both cases the Nielsen identity is a combination of a quantum
equation of motion, the Dyson-Schwinger equation, and the Slavnov-Taylor identity (STI) or
diffeomorphism constraint. The setup also entails that correlations of
the fluctuation fields are necessarily background-dependent. This is
easily seen by iterating \eqref{eq:NI}. Moreover, in the linear split,
diffeomorphism invariance of the observables is encoded in non-trivial
STIs for the fluctuation correlation
functions, while in the geometrical formulation, the non-trivial STIs
are encoded in expectation values of $f(\bar g,h)$ and its
derivatives. 

Due to \eqref{eq:NI} we have to deal with the peculiarity that background
independence and physical diffeomorphism invariance of observables
necessitate background-dependence and non-trivial STIs for the
correlation functions of the fluctuation fields. This leads to
seemingly self-contradictory statements: in particular, for
the quantum effective action $\Gamma[\bar g, h]$ it entails that physical
diffeomorphism invariance of observables is not achieved by
diffeomorphism invariance w.r.t.\ diffeomorphism transformations of
the fluctuation fields. The latter does not do justice to either 
diffeomorphism invariance or background independence.

This peculiarity can easily be checked in a non-Abelian gauge theory
within the background field formulation: in a fluctuation gauge
invariant approximation to the effective action, even two-loop
universal observables such as the two-loop $\beta$-function cannot be
computed correctly. Indeed, in this case it is well-known that only
the non-trivial STIs for the fluctuation gauge field elevate the
auxiliary background gauge invariance to the physical one holding for
observables, see e.g.~\cite{DeWitt:1967ub,Abbott:1981ke}. 

The above considerations underline the importance of a direct
computation of correlation functions of the fluctuation field
$h$. Indeed, the corresponding set of flow equations for
$\Gamma^{(n)}=\Gamma^{(0,n)}$ is closed in the sense that the flow
diagrams only depend on $\Gamma^{(n)}$ with $n\geq 2$.  Here,
$\Gamma^{(n,m)}$ stands for the $n$-th background field derivative and
$m$-th fluctuation field derivative of the effective action,
\begin{align}\label{eq:nm}
  \Gamma^{(n,m)}[\bar g,\phi] \definition \0{\delta^{n+m}\Gamma[\bar
    g,h]}{\delta \bar g^n \delta \phi^m}\,.
\end{align}
In turn, the flows for pure
background, $\Gamma^{(n,0)}$, or mixed background-fluctuation
functions, $\Gamma^{(n,m)}$ with $m\neq 0$, necessitate the fluctuation
correlation functions as an input: the background correlation
functions can be iteratively computed in powers of the background
metric. In other words, the dynamics of the system is solely
determined by the pure fluctuation correlation functions.

For this reason, several approaches for computing correlation
functions of the fluctuation field have been put forward in the last
years.  Some of these approaches are set up for computing both
correlation functions of the fluctuation field as well as those of the
background field: vertex expansion
\cite{Christiansen:2012rx,Codello:2013fpa,Christiansen:2014raa,
  Christiansen:2015rva,Meibohm:2015twa,Meibohm:2016mkp,Eichhorn:2016esv}
and bimetric approach
\cite{Manrique:2009uh,Manrique:2010mq,Manrique:2010am,Becker:2014qya}.
Another set of approaches relies on utilising the Nielsen or split
Ward identities explicitly or implicitly,
\cite{Pawlowski:2003sk,Donkin:2012ud,Dietz:2015owa,Safari:2015dva,
  Morris:2016nda,Labus:2016lkh,Safari:2016gtj,Safari:2016dwj,
  Wetterich:2016ewc,Morris:2016spn,Percacci:2016arh}.

So far, the $\Gamma^{(n)}$ for $n\geq 2$ have been computed directly
in only one approach, the vertex expansion, see
\cite{Christiansen:2012rx,Codello:2013fpa,Christiansen:2014raa,Christiansen:2015rva}
for pure gravity and
\cite{Meibohm:2015twa,Meibohm:2016mkp,Eichhorn:2016esv} for
gravity-matter systems. A mixture of vertex expansion and background
approximation has been used in
\cite{Codello:2013fpa,Dona:2013qba,Dona:2015tnf}. Present results
include $\Gamma^{(n)}$ for $n=0,1,2,3$, where higher vertices have
been estimated by lower ones, relying on approximate covariance of the
correlation functions. Such a structure has already been confirmed in
the perturbative and semi-perturbative regime of QCD, see
\cite{Mitter:2014wpa}.

\section{Effective action \& Functional Renormalisation Group}
\label{sec:frg}
The set of (covariant) correlation functions of the metric, $\langle
g(x_1)\cdots g(x_n)\rangle $, defines a given theory of quantum
gravity. All observables can be constructed from these basic building
blocks. The correlation functions are generated from the single metric
effective action, $\Gamma[g]=\Gamma[g, h=0]$, which is the free energy
in a given metric background $g=\bar g +h$ at $h=0$. Here we have
restricted ourselves to a linear split. The underlying classical
action is the Einstein-Hilbert action,
\begin{align}\label{eq:EH_action}
  \actionCl_\text{EH} &= \frac{1}{16 \pi \newtonG} \int \dn{4}{x} \sqrt{\det g}
  \Bigl(2\cosmologicalConstant-\R(g)\Bigr) + \actionGF + \actionGhost
  \,,
\end{align}
where $\R(g)$ is the Ricci curvature scalar, while $\actionGF[\bar
g,h]$ and $\actionGhost[\bar g, \phi]$ describe the gauge-fixing and
Faddeev-Popov ghost parts of the action, respectively. The gauge fixing
action reads
\begin{align} \label{eq:gf-action}
  \actionGF[\bar g,h]=\frac1{2\alpha}\int\mathrm d^4x\,
  \sqrt{\det\bar g}\,\bar g^{\mu\nu}F_\mu F_\nu\,.
\end{align}
We employ a linear, de-Donder type gauge-fixing, 
\begin{align} \label{eq:gf-condition} F_\mu &= \bar \nabla^\nu
  h_{\mu\nu} - \frac{1+\beta}4 \bar \nabla_\mu {h^\nu}_\nu \,.
\end{align}
In particular we use the harmonic gauge given by $\beta=1$.  This
choice simplifies computations considerably due to the fact that the
poles of all modes of the classical graviton propagator coincide.  We
furthermore work in the limit of a vanishing gauge parameter, $\alpha \to
0$. This is favourable because then the gauge does not change during
the flow since $\alpha=0$ is a RG fixed point \cite{Litim:1998qi}.
Moreover, it allows for a clear separation of propagating and
non-propagating degrees of freedom. The ghost part of the action reads 
\begin{align}\label{eq:gh-action}
  \actionGhost[\bar g,\phi]=\int\mathrm d^4x \,\sqrt{\det\bar{g}}\, \bar{c}^\mu
  \FPoperator_{\mu\nu} c^\nu \,,
\end{align}
where $\bar c$ and $c$ denote the (anti-) ghost field and
$\FPoperator$ is the Faddeev-Popov operator deduced from
\eqref{eq:gf-condition}. For $\beta=1$ it is given by 
\begin{align}\label{eq:gh-faddeev-popov}
 \FPoperator_{\mu\nu} &=
  \bar\nabla^\rho\left(g_{\mu\nu}\nabla_\rho+g_{\rho\nu}\nabla_\mu
  \right)-\bar\nabla_\mu\nabla_\nu\,.
\end{align}
The gauge fixing and ghost term in \eqref{eq:gf-action} 
and \eqref{eq:gh-action} introduce the separate dependence on $\bar g$ 
and $h$ leading to the non-trivial Nielsen identities in \eqref{eq:NI}.

\subsection{Flow equation}
\label{sec:flow}
An efficient way of computing non-perturbative correlation functions
is the functional renormalisation group. In its form for the effective
action, see \cite{Wetterich:1992yh, Ellwanger:1993mw,Morris:1993qb},
it has been applied to quantum gravity \cite{Reuter:1996cp}. For
reviews on the FRG approach to gauge theories and gravity see 
e.g.~\cite{Pawlowski:2005xe,Gies:2006wv,Niedermaier:2006wt,Percacci:2007sz,Litim:2011cp,Reuter:2012id}.
The RG flow of the effective action for pure quantum gravity is given
by
\begin{align} \label{eq:gen_flow_eq} {\scaleDerivative{\actionQu}}_k
  &= \frac{1}{2} \tr\left[
    \frac{1}{\actionEff{k}^{(2)}+\regulator_{k}}
    \scaleDerivative{\regulator}_k
  \right]_{\graviton\graviton}\hspace*{-0.35cm} - \tr\left[
    \frac{1}{\actionEff{k}^{(2)}+\regulator_{k}}
    \scaleDerivative{\regulator}_k
  \right]_{\antighost\ghost}\hspace*{-0.2cm}.
\end{align}
Here, $\scaleDerivative{} \equiv k \partial_k$ denotes the scale
derivative, where $k$ is the infrared cutoff scale.
$\actionEff{k}^{(2)}=\Gamma_k^{(0,2)}$ stands for the second
fluctuation field derivative of the effective action, while
$\regulator_k$ is the regulator, which suppresses momenta below $k$.
The trace in \eqref{eq:gen_flow_eq} sums over internal indices and integrates over
space-time.

The introduction of cutoff terms leads to regulator-dependent
modifications of STIs and NIs that vanish for $R_k\to 0$. The
respective symmetry identities have hence been named modified
Slavnov-Taylor identities (mSTIs) and modified Nielsen- or split Ward
identities (mNIs/mSWIs). The modification entails the breaking of the
physical or quantum diffeomorphism invariance in the presence of a
background covariant momentum cutoff. Still, background diffeomorphism
invariance is maintained in the presence of the cutoff term.

\subsection{Covariant expansion}\label{sec:covexp}
The effective action $\actionEff{k} [\metricBackground,
\superfieldFluctuating]$ depends on the background metric
$\metricBackground$ and the fluctuation superfield
$\superfieldFluctuating=(h,c,\bar c)$, see \eqref{eq:phi},
separately. The functional flow equation \eqref{eq:gen_flow_eq} is
accompanied by the functional mSTIs \& mNIs for the effective action
that monitor the breaking of quantum diffeomorphism invariance,
see~\eqref{eq:NI} in \autoref{sec:Diff}.

In order to solve \eqref{eq:gen_flow_eq}, we employ a vertex
expansion around a given background $\metricBackground$, to wit
\begin{align} \label{eq:vertex_expansion_qeg}
  \actionQu_k\left[\metricBackground,\superfieldFluctuating\right] =
  \sum_{n=0}^{\infty} \frac{1}{n !}
  \actionQu_k^{(\phi_{i_1}\ldots\phi_{i_n})}\left[\metricBackground,0\right]\,
  \prod_{l=1}^n \phi_{i_l} \;,
\end{align}
where the superscript fields in parentheses are a short-hand notation
for field derivatives, and where contracting over super-indices $i_j$
occurring twice is implied.  In this work, we include the full flow of
the vertex functions up to the graviton four-point function.

As discussed in \autoref{sec:Diff}, the expansion coefficients
$\actionQu_k^{(\phi_{i_1}\ldots\phi_{i_n})}=\Gamma_k^{(n)}=\Gamma_k^{(0,n)}$
satisfy mSTIs as well as mNIs with $\Gamma^{(n,m)}_k$ being defined in
\eqref{eq:nm}. For the sake of simplicity we now restrict ourselves to
the gauge fixing used in the present work, \eqref{eq:gf-condition} with
$\alpha=0$. Then the fluctuation graviton propagator is transverse: it
is annihilated by the gauge fixing condition.

An important feature of the functional RG equations is that for
$\alpha=0$ the flow equations for the transverse vertices
$\Gamma^{(n)}_{k,T}$ are closed: the external legs of the vertices in
the flow are transverse due to the transverse projection of the flow,
the internal legs are transverse as they are contracted with the
transverse propagator. Schematically this reads
 \begin{align}\label{eq:flowbot}
\partial_t \Gamma^{(n)}_{k,T}={\flow}^{(n)}_T[\{\Gamma^{(m)}_{k,T}\}]\,. 
\end{align}
In other words, the system of transverse fluctuation correlation
functions is closed and determines the dynamics of the system.  On the
other hand, the mSTIs are non-trivial relations for the longitudinal
parts of vertices in terms of transverse vertices and
longitudinal ones.  This leads us to the schematic relation
\begin{align}\label{eq:mSTI}
\Gamma^{(n)}_{k,L}=\mathrm{mSTI}^{(n)}[\{\Gamma^{(m)}_{k,T}\}, \{\Gamma_{k,L}^{(m)}\}]\,, 
\end{align}
see \cite{Fischer:2008uz} for non-Abelian gauge theories.  In
consequence, the mSTIs provide no direct information about the
transverse correlation functions without further constraint. In the
perturbative regime this additional constraint is given by the
uniformity of the vertices, for a detailed discussion in non-Abelian
gauge theories see \cite{Cyrol:2016tym}.

Accordingly, our task reduces to the evaluation of the coupled set of
flow equations for the transverse vertices $\Gamma_{k,T }^{(n)}$.  Each
transverse vertex can be parameterised by a set of
diffeomorphism-invariant expressions.  Restricting ourselves to local
invariants and second order in the curvature we are left with
\begin{align}\label{eq:diffinv}
\R\,,\quad \R^2\,,\quad \R_{\mu\nu}^2\,.
\end{align} 
The square of the Weyl tensor $C^2$ is eliminated via the Gauß-Bonnet 
term, which is a topological invariant. Higher-derivative terms, such as 
\begin{align}\label{eq:highder}
R^{\mu\nu} f_{\mu\nu\rho\sigma}(\nabla)R^{\rho\sigma}\qquad \text{with}\qquad  f(0)= 0\,,
\end{align}
are also taken into account. Without the constraint $f(0)=0$, equation
\eqref{eq:highder} also includes $R^2$ and $\R_{\mu\nu}^2$, more details
on this basis can be found in \autoref{sec:flows}. Note that also
non-diffeomorphism-invariant terms are generated by the flow.  In
\autoref{sec:mom-dep-nptfct} we discuss all invariants which are
included in the parameterisation of our vertices.

For the background vertices $\Gamma_k^{(n,0)}$ we use the following:
the NIs become trivial in the IR as we approach classical gravity, as
shown in \autoref{sec:IR-behaviour}. Moreover, for one of the two IR
fixed points this implies that the derivative with respect to a
background field is the same as a derivative with respect to a
fluctuation field. This allows us to impose the trivial NIs in the IR,
and all couplings are related. Then, the couplings at $k> 0$ follow
from the flow equation. However, for the fluctuation couplings this
amounts to solving a fine-tuning problem in the UV, for more details
see \autoref{sec:IR-behaviour}. The latter is deferred to future work.

\section{Flows of Correlation Functions}\label{sec:flows}
In this chapter we discuss the technical details of the covariant
expansion scheme used in the present work, including the approximations
used and their legitimisation. In our opinion, a careful reading of
this chapter is essential for a full understanding of the results
obtained in the present work. This applies in particular to
\autoref{sec:invariants}.

\subsection{Covariant tensors and uniformity} \label{sec:invariants}
\begin{figure}[!]
  \includegraphics[width=\linewidth]{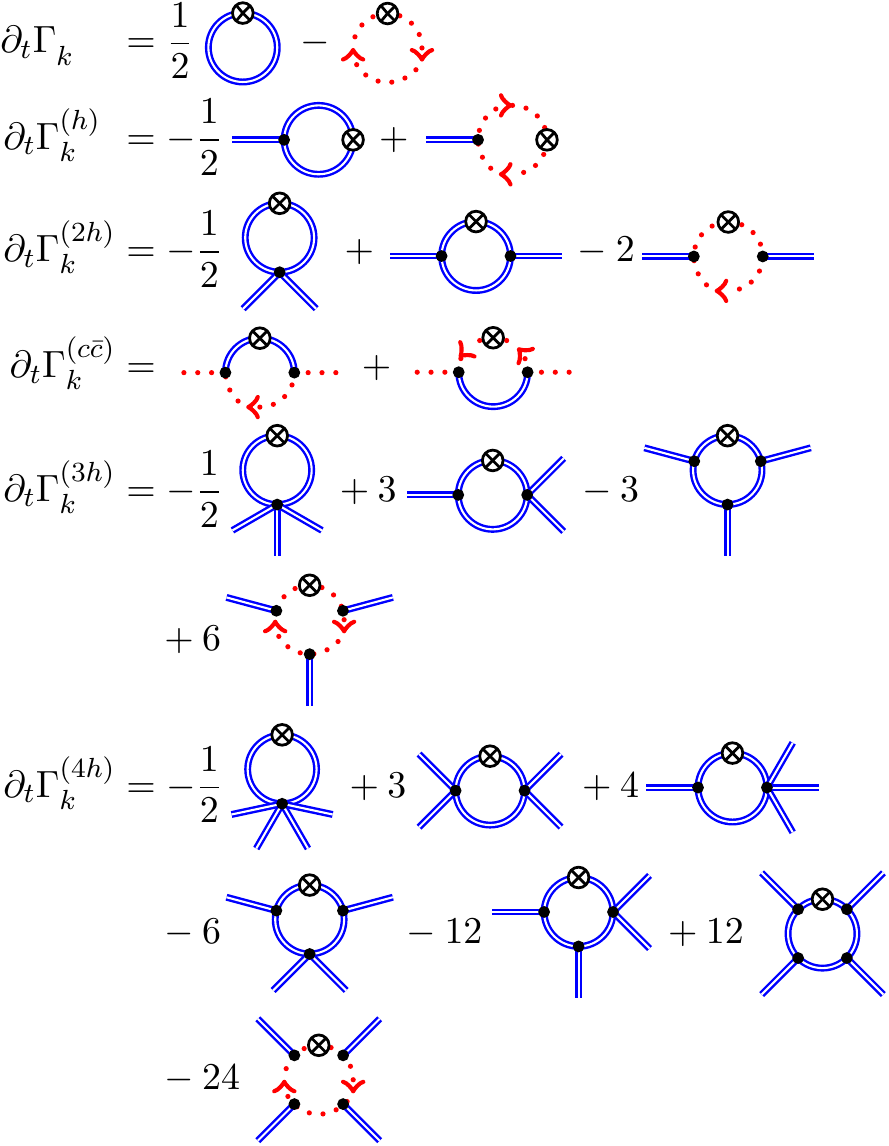}
  \caption{Diagrammatic representation of the flow of the vertex
    functions up to the graviton four-point function. The flow of any
    $n$-point function depends on the $(n+1)$- and $(n+2)$-point
    functions. Double and dotted lines represent graviton and ghost
    propagators, respectively.  All vertices are dressed and denoted
    by filled circles. Crossed circles stand for regulator
    insertions. Symmetrisation with respect to interchange of
    external momenta $p_i$ is understood.}
  \label{fig:flow_n_point}
\end{figure}

The flows of the $n$-point correlation functions are generated from
the FRG equation \eqref{eq:gen_flow_eq} by taking $n$-th order fluctuation
field derivatives in a background $\bar g$, (see
\autoref{fig:flow_n_point}).  In order to solve the flow equation, we
employ a vertex ansatz~\cite{Christiansen:2012rx,Fischer:2009tn}
including the flow of all relevant vertices up to the graviton
four-point function.  This vertex ansatz disentangles the couplings of
background and fluctuation fields by introducing individual couplings
$\Lambdan{n}$ and $\Gn{n}$ for each \nptfct{n}.  These individual
couplings are introduced at the level of the $n$-point correlators and 
replace the cosmological constant $\cosmologicalConstant$ and Newton's
coupling $\newtonG$ of the classical Einstein-Hilbert action after performing the
respective field derivatives.  In summary, for
the flat background $\bar g=\delta$ our vertex ansatz reads
\begin{align}\label{eq:vertex}
  \actionEff{k}^{(\phi_1 \mathellipsis \phi_n)}(\vect{p}) =
  \left(\prod_{i=1}^n \wfr[\frac{1}{2}]{\phi_i}{p_i^2}\right)
  \Gn{n}^{\frac{n}{2}-1}(\vect{p}) \tensorStructure^{(\phi_1 \mathellipsis
    \phi_n)}(\vect{p};\Lambdan{n}) \,,
\end{align}
where
\begin{align}
  \label{eq:tensor_struc}
  \mathcal{T}^{(\phi_1 \mathellipsis \phi_n)}(\vect{p};\Lambdan{n})&=
  \newtonG\,\actionCl_\text{EH}^{(\phi_1 \mathellipsis
    \phi_n)}(\vect{p};\cosmologicalConstant \rightarrow \Lambdan{n})\,,
\end{align}
denote the tensor structures extracted from the classical gauge-fixed
Einstein-Hilbert action \eqref{eq:EH_action}. The only flowing
parameter in these tensors $\mathcal{T}^{(\phi_1 \mathellipsis
  \phi_n)}$ is $\Lambda_n$, while $G_n(\vect{p})$ carries the global
scale- and momentum dependence of the vertex. In the above equations,
$\vect{p} = (p_{\phi_1},\mathellipsis,p_{\phi_n})$ denotes the momenta
of the external fields $\phi_i$ of the vertex.

Apart from their flow equations, the $n$-point functions in
\eqref{eq:vertex} also satisfy standard RG-equations, see 
e.g.~\cite{Pawlowski:2005xe}. These RG-equations entail the
reparameterisation invariance of the theory under a complete rescaling
of all scales including $k$. With the parameterisation given in
\eqref{eq:vertex}, this RG-running is completely carried by the wave
function renormalisations $\wfr{\phi_i}{p_i^2}$ of the fields
$\phi_i$, see e.g.~\cite{Fischer:2009tn,Christiansen:2014raa,Meibohm:2015twa}. 
Consequently, the $G_n$
and $\Lambda_n$ are RG-invariant, and hence are more directly related
to observables such as $S$-matrix elements. This parameterisation of
the vertices also ensures that the wave function renormalisations
never appear directly in the flow equations, but only via the
anomalous dimensions
\begin{align}
  \anomalousDimension{\phi_i} (p_i^2) \definition - 
\scaleDerivative{\ln \wfrsymb}_{\phi_i}(p_i^2)\,.
\end{align}

$\Gn{n}(\vect{p})$ is the gravitational coupling of the $n$-point
function, while $\Lambdan{n}$ denotes the momentum-independent part of
the correlation function.  In particular, $\Lambdan{2}$ is related to
the graviton mass parameter $\Mass^2 \definition -2 \Lambdan{2}$. Finally, 
all the parameters $\wfrsymb_{\phi_i}$, $\Gn{n}$, and $\Lambdan{n}$
are scale-dependent, but we have dropped the subscript $k$ in order to 
improve readability.

In principle, all tensor structures, including
non-diffeomorphism-invariant ones, are generated by the flow, but for
our vertex functions we choose to concentrate on the classical
Einstein-Hilbert tensor structures in the presence of a non-vanishing
cosmological constant. Despite the restriction to these tensor
structures, the \nptfct{n}s have an overlap with higher curvature invariants via the
momentum dependence of the gravitational couplings. For example,
the complete set of invariants that span the graviton wave function 
renormalisation is given by 
\begin{align}\label{eq:Rf}
\R\,,\qquad  \RicciTensor^{\mu\nu}
f^{(2)}_{\mu\nu\rho\sigma}(\nabla) \RicciTensor^{\rho\sigma}\,, 
\end{align}
where the superscript indicates that is a covariant tensor
contributing to the two-point correlation function.  Note also that we now
drop the restriction on $f$ present in \eqref{eq:highder}. Then, this
invariant naturally includes $R^2$ and $R_{\mu\nu}^2$ as the lowest
order local terms. If we also allow for general momentum-dependencies,
the corresponding covariant functions $f$ are given by given by
\begin{align}\nonumber 
  f^{(2)}_{ R^2,\mu\nu\rho\sigma} =
  &\,\delta_{\mu\nu}\delta_{\rho\sigma} P^{(2)}_{R^2}(-\nabla^2)\,,\\[2ex] 
  f^{(2)}_{R_{\mu\nu}^2,\mu\nu\rho\sigma} = &\,\012 \left(
    \delta_{\mu\rho}\delta_{\nu\sigma}+\delta_{\mu\sigma}\delta_{\nu\rho}\right)
  P^{(2)}_{R_{\mu\nu}^2}(-\nabla^2)\,.
\label{eq:fRR2}
\end{align}
The lowest order local terms, $R^2$ and $R_{\mu\nu}^2$, are given by
$P^{(2)}_{R^2}=1$ and $P^{(2)}_{R_{\mu\nu}^2}=1$, respectively. Note
that \eqref{eq:fRR2} also allows for non-local terms in the
IR, i.e.\ anomaly-driven terms with $P^{(2)}_{R^2}=1/\nabla^2$, see
e.g.~\cite{Mottola:2010gp}.  In turn, higher curvature invariants do
not belong to the set of the graviton wave function renormalisation
since they are at least cubic in the graviton fluctuation field.

In the present work we resort to a uniform graviton propagator in
order to limit the already large computer-algebraic effort involved.
The uniform wave function renormalisation is then set to be that of
the combinatorially dominant tensor structure, the
transverse-traceless graviton wave function renormalisation, thereby
estimating the wave function renormalisations of the other modes by
the transverse-traceless one. Such uniform approximations have been
very successfully used in thermal field theory. There, usually the
tensor structures transverse to the heat-bath are used as the uniform
tensor structure, for a detailed discussion see e.g.~\cite{Schnoerr:2013bk} 
and references therein. This approximation is
typically supported by combinatorial dominance of this tensor
structure in the flow diagrams.  Indeed, as already indicated above,
the transverse-traceless mode gives the combinatorially largest
contribution to the flow of the vertices computed here.  Note that
such an approximation would get further support if the $R$ tensor
structures dominate the flows, which indeed happens in the present
computation.

Within this approximation the $\R^2$ tensor structures drop out on the
left-hand side of the graviton flow, since $\R$ is already quadratic
in the transverse-traceless graviton fluctuation field: in other
words, the tensors defined by $f^{(2)}_{R^2}$ in \eqref{eq:fRR2} have
no overlap with the transverse-traceless graviton.

The set of invariants that span the gravitational coupling
$\Gn{3}(\vect{p})$ is given by 
\begin{align} \label{eq:g3-tensors}
\R\,, \qquad \RicciTensor^{\mu\nu}
f^{(3)}_{\mu\nu\rho\sigma}(\nabla) \RicciTensor^{\rho\sigma}\,, \qquad 
\RicciTensor^{\mu\nu} \RicciTensor^{\rho\sigma}
f^{(3)}_{\mu\nu\rho\sigma\omega\zeta}(\nabla) \RicciTensor^{\omega\zeta}\,. 
\end{align}
Again, the invariants $\R^2$ and $\R^3$ can be excluded from this set
due to their order in transverse-traceless graviton fluctuation
fields. In consequence, $\Gn{4}(\vect{p})$ is the only coupling in our 
setup that has overlap with $\R^2$ contributions and higher terms in $f^{(4)}_{R^2}$.

Furthermore, in \autoref{sec:mom-dep-nptfct} we show that the by far
dominant contribution to $\Gn{3}(\vect{p})$ in the momentum range
$0\leq p^2\leq k^2$ stems from the invariant $\R$.  All higher
momentum dependencies of the graviton three-point function are covered
by the momentum dependence of the graviton wave function
renormalisation.  This was already observed
in~\cite{Christiansen:2015rva}. As already briefly mentioned above, it
gives further support to the current uniform approximation: the
assumption of uniformity allows us to restrict ourselves to computing
the Einstein-Hilbert tensor structure for the transverse-traceless
graviton as the combinatorially dominating tensor structure. The
striking momentum-independence of the actual numerical flows supports
a momentum-independent approximation of $\Gn{3}$.  In terms of
\eqref{eq:fRR2} it implies that the dominant tensor structure
for the transverse-traceless mode is given by $f^{(3)}_R$ with
$P^{(3)}_R=1$. The $R_{\mu\nu}^2$ tensor structure vanishes
approximately, see~\eqref{eq:f3Ricci=0}.

In contrast to the situation for the two- and three point function,
the $R^2$ invariant overlaps with our transverse-traceless projection
for the graviton four-point function. Indeed, its flow receives
significant contributions from the invariant $\R^2$. It follows that
for the graviton four-point function $\R$ is not the only dominant
invariant in the momentum range from $p=0$ to $p=k$, as we show in
\autoref{sec:mom-dep-nptfct}. In consequence we either have to
disentangle contributions from $\R$ and $\R^2$ tensor structures in
terms of an additional tensor structure or we resolve the momentum
dependence of $\Gn{4}(\vect{p})$. In the present work we follow the
latter procedure, see \autoref{sec:flows-couplings} for details.

\begin{figure*}[tbp]
  \centering
  \includegraphics[width = \textwidth]{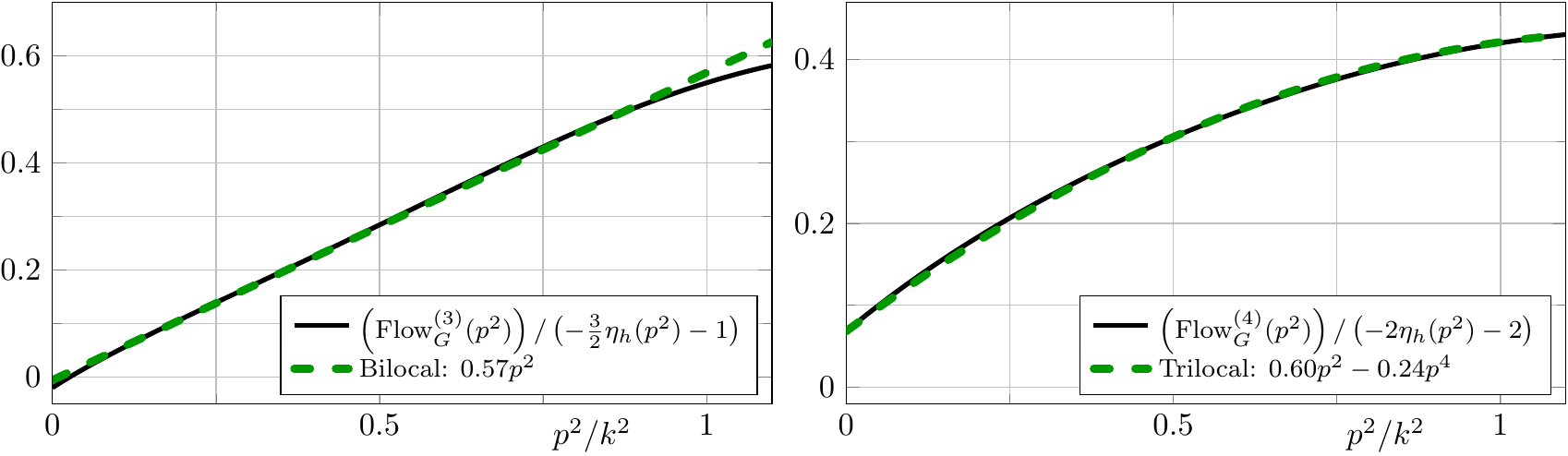}
  \caption[Momentum dependence of the flow of the graviton
  \nptfct{n}]{ Momentum dependence of the flow of the graviton
    three-point function (left) and the graviton four-point function
    (right) divided by $(-\frac{n}{2}
    \anomalousDimension{\graviton}(p^2) - n + 2)$ as defined in
    \eqref{eq:flow_polynomial}.  The flows are evaluated at
    $\left( \mu, \lambdan{3}, \lambdan{4}, \gn{3}, \gn{4} \right) =
    \left( -0.4, 0.1, -0.1, 0.7, 0.5 \right)$ and
    $\lambdan{6}=\lambdan{5}=\lambdan{3}$ as well as
    $\gn{6}=\gn{5}=\gn{4}$.  The flows have such a simple polynomial
    structure as long as all couplings $\lambdan{n}$ remain small,
    i.e.\ $\lvert\lambdan{n}\rvert\lesssim1$. Importantly, the inclusion 
    of a $p^4$ term in the left panel offers no significant improvement. 
    Note that the constant parts of
    the functions are irrelevant for the beta functions since they are
    extracted from a different tensor projection. For $p^2>k^2$ the 
    momentum dependence of the flows is not polynomial anymore. 
    }
\label{fig:mom_dep_flow}
\end{figure*}

\subsection{Projection onto \texorpdfstring{\nptfct{n}s}{n-point
    functions}} \label{sec:flows-proj}
The flow equations for the couplings $\Lambdan{n}$ and $\Gn{n}$ are
obtained by the following projection onto the flow of the graviton
$n$-point functions $\scaleDerivative{\actionQu}_k^{(n)}$. We use the
classical Einstein-Hilbert tensor structures
$\tensorStructure^{(n)}(\vect{p};\Lambdan{n})$ as a basis for our
projection operators.  Furthermore, we project onto the spin-two
transverse-traceless part of the flow, which is numerically
dominant. Moreover, it does not depend on the gauge. This
transverse-traceless projection operator is then applied to all
external graviton legs. The flow of the couplings $\Lambdan{n}$ is
then extracted with the help of the momentum-independent part of said
tensor structures, namely $\projLambdan{n} \definition
\tensorStructure^{(n)}(0;\Lambdan{n})/\Lambdan{n}$.  For the couplings
$\Gn{n}$ we use $\projGn{n}\definition
\tensorStructure^{(n)}(\vect{p};0)/p^2$. Dividing by $\Lambdan{n}$
and $p^2$ ensures that the projection operators are dimensionless and
scale-independent. 

In principle, the flow of any $n$-point function depends on all
external momenta $p_i, i\in \{1,\mathellipsis,n\}$, where e.g.\ $p_n$
can be eliminated due to momentum conservation.  For the two-point
function, the momentum configuration is trivial, and only one
momentum squared, $p^2$, needs to be taken into account.  In contrast, this
dependence becomes increasingly complex for the higher $n$-point
functions: The three-point function depends on three parameters (two
momenta squared and one angle), the four-point function already depends on six
parameters, and so on. To simplify the computations, we use a
maximally symmetric $(n-1)$-simplex configuration for all
$n$-point-functions, thereby reducing the momentum dependence to a
single scalar parameter.  This symmetric momentum configuration was
already used for the graviton three-point function
in~\cite{Christiansen:2015rva}.  In the context of Yang-Mills
theories, this approximation has been shown to be in good agreement
with lattice computations on the level of the flow of the propagator
\cite{Cyrol:2016tym}.  Notably, in the symmetric momentum
configuration all external momenta have the same absolute value $p$,
and the same angles between each other.  The scalar product of any two
momenta in this momentum configuration then reads
\begin{align}
  p_i \cdot p_j = \frac{n \delta_{i j} - 1}{n-1} p^2\,,
\end{align}
where $\delta_{i j}$ denotes the Kronecker delta.  Note that such a
symmetric momentum configuration only exists up to the \nptfct{(d+1)},
where $d$ is the dimension of spacetime.

In the following, the expressions $\flow^{(n)}$ stand for the
dimensionless right-hand sides of the flow equations divided by
appropriate powers of the wave function renormalisations.  More
explicitly, we define
\begin{align} \label{eq:def-flow} \flow^{(n)}_i(p^2) \definition
  \frac{\scaleDerivative{ \actionQu}^{(n)}_{i}(p^2)
  }{\wfr[\frac{n}{2}]{\phi}{p^2} k^{2-n}}\,,
\end{align}
where the index $i$ represents the projection on some
tensor structure.  In this work, we use the transverse-traceless
projection operator $\projTT$, the projection operators
$\projGn{n}$ and $\projLambdan{n}$ mentioned earlier for the graviton
\nptfct{n}s, as well as the transverse projection operator $\proj_\tSymb$
for the ghost propagator.  Note that the objects $\flow^{(n)}_i$ do
not contain any explicit factors of the wave function renormalisations
$\wfrsymb_{\phi}$.  Instead, their running appears via the anomalous
dimensions $\anomalousDimension{\phi}$.

Last but not least, we choose to model the regulator functions
$\regulatorOfField{\phi_i}$ on the corresponding two-point functions
at vanishing mass, i.e.
\begin{align} \label{eq:regulator_construction}
  \regulatorOfField{\phi_i} (p_i^2) = \left.
\actionQu^{(\phi_i \phi_i)} (p_i^2) \right|_{m_{\phi_i} = 0} \shapeFunction_{\phi_i} ( p_i^2/k^2 ) \,.
\end{align}
Here, $\shapeFunction_{\phi_i} ( p_i^2/k^2 )$ denotes the regulator
shape function.  For all fields in this work, we choose the Litim
regulator \cite{Litim:2000ci}, to wit
\begin{align} \label{eq:shape_function_sharp}
  \shapeFunction ( x ) = \left( x^{-1} - 1 \right) \ThetaFunction\left( 1 - x \right) \,.
\end{align}
This choice allows for analytic flow equations for all couplings that
are evaluated at vanishing external momenta.

Furthermore, we introduce the dimensionless couplings
\begin{align} \label{eq:couplings_dimless}
  \mass \definition \Mass^2 k^{-2}, \quad \lambdan{n} \definition 
\Lambdan{n} k^{-2}, \quad \gn{n} \definition \Gn{n} k^2 \,.
\end{align}
At the UV and IR fixed points, the flow of these dimensionless couplings vanishes.

\subsection{Momentum dependence of the graviton
  \texorpdfstring{\nptfct{n}s}{n-point
    functions}} \label{sec:mom-dep-nptfct}
We now investigate the momentum dependence of the
flow of the graviton \nptfct{n}s as defined in \eqref{eq:def-flow}.
We restrict ourselves to the momentum range $0
\leq p^2 \leq k^2$ as well as to the transverse-traceless part of the
graviton \nptfct{n}s.

The first non-trivial result is that the flows of the graviton three-
and four-point functions
projected on the tensor structure of the gravitational coupling and
divided by $(-\frac{n}{2} \anomalousDimension{\graviton}(p^2) - n +
2)$ are well described by a polynomial in $p^2$, provided that the
couplings $\lambdan{n}$ are small, i.e.
\begin{align} \label{eq:flow_polynomial} \cfrac{\flow_G^{(3)}
    (p^2)}{-\frac{3}{2} \anomalousDimension{
      \graviton}(p^2) - 1} &\approx a_{0} + a_{1}\, p^2 \nonumber \,,\\[2ex]
  \cfrac{\flow_G^{(4)} (p^2)}{-2 \anomalousDimension{\graviton}(p^2) -
    2} &\approx b_{0} + b_{1}\, p^2 + b_{2}\, p^4 \,,
\end{align}
with some constants $a_i$ and $b_i$ that depend on the evaluation point in theory
space. This momentum dependence is displayed in
\autoref{fig:mom_dep_flow}. We emphasise that these equations only
hold in the momentum range $0 \leq p^2 \leq k^2$, if the flow is
generated by Einstein-Hilbert vertices, and if the constant parts of
the vertices are small, i.e.\ $\lvert\lambdan{n}\rvert \lesssim 1$.  If the
condition of small $\lambdan{n}$ is violated, then the flow as in
\eqref{eq:flow_polynomial} is non-polynomial.  We did not
compute the flow generated by an action including higher
curvature terms, however, we suspect that the flow will still be
polynomial but possibly of a higher degree. 

It is important to note that the graviton three- and four-point
functions have a different highest power in $p^2$.  This is a second
non-trivial result for the following reasons: as already mentioned
before, the coupling $\gn{3}(p^2)$ has an overlap with $\R$ and
$\RicciTensor_{\mu\nu}^2$, and higher derivative terms in
$f^{(3)}_{\RicciTensor_{\mu\nu}^2}$, but not with any $\R^2$ tensor
structures in $f^{(3)}_{R^2}$, c.f.\ \eqref{eq:fRR2}. For
example, the generation of $\RicciTensor_{\mu\nu}^2$ with
$P^{(3)}_{\RicciTensor_{\mu\nu}^2}=1$ would manifest itself in a
$p^4$-contribution to the flow of the graviton three-point function.
Equation \eqref{eq:flow_polynomial} and \autoref{fig:mom_dep_flow}
show that such a $p^4$-contribution as well as higher ones are
approximately vanishing. This demonstrates in particular that the
generation of $\RicciTensor_{\mu\nu}^2$ is non-trivially
suppressed. In other words,
\begin{align}\label{eq:f3Ricci=0}
  f^{(3)}_{\RicciTensor_{\mu\nu}^2}\approx 0\,,
\end{align}
where the superscript indicates the three-graviton vertex.

On the other hand, the projection on $\gn{4}(p^2)$ overlaps with $\R$,
$\RicciTensor_{\mu\nu}^2$, $\R^2$ tensor structures, and the related
higher derivatives terms in $f^{(4)}_{\RicciTensor_{\mu\nu}^2}$ and
$f^{(4)}_{R^2}$. It also overlaps with curvature invariants to the third
power with covariant tensors such as $f^{(4)}_{R_{\mu\nu}^3}$ and similar
ones. Note that it has no overlap with $f^{(4)}_{R^3}$.

Similarly to possible $p^4$-contributions for the three-graviton
vertex, $p^6$-contributions and even higher powers in $p^2$ could be
generated but are non-trivially suppressed.  The $p^4$-contribution to
the flow, which is described in \eqref{eq:flow_polynomial}
and displayed in \autoref{fig:mom_dep_flow}, could stem from either
$\R^2$ or $\RicciTensor_{\mu\nu}^2$ tensor structures. Now we use
\eqref{eq:f3Ricci=0}.  It entails that the graviton three-point vertex
does not generate the diffeomorphism invariant term $R_{\mu\nu}^2$
although it has an overlap with it.  This excludes $R_{\mu\nu}^2$ as a
relevant UV direction, which would otherwise be generated in all
vertices.  This statement only holds if we exclude non-trivial
cancellations of which we have not seen any signature. Accordingly we
set
\begin{align}\label{eq:f4Ricci=0}
  f^{(4)}_{\RicciTensor_{\mu\nu}^2}\approx 0\,,
\end{align}
and conclude that this $p^4$-contribution or at least its UV-relevant
part stems solely from $\R^2$. It may be used to determine
$f^{(4)}_{R^2}$.

In summary, the above statements about the momentum-dependencies
are highly non-trivial and show that $R^2$-contributions are
generated while $R_{\mu\nu}^2$ and other higher derivative terms are
strongly suppressed. These non-trivial findings also allow us to
determine the most efficient way to project precisely onto the
couplings of different invariants. This is discussed in
\autoref{sec:flows-couplings}.

We close this section with a brief discussion of the effect of higher
derivative terms on perturbative renormalisability and the potential
generation of massive ghost states.  As already discussed
in~\cite{Stelle:1977ry} in a perturbative setup, it is precisely the
$R_{\mu\nu}^2$ term which makes the theory perturbatively
renormalisable. However, in this setup it gives rise to negative norm
states.  On the other hand, the $R^2$ term neither ensures
perturbative renormalisability, nor does it generate negative norm
states.  This is linked to the fact that the $R^2$ term does not
contribute to the transverse traceless part of the graviton
propagator.  Consequently, the non-trivial suppression of
$R_{\mu\nu}^2$ tensor structures might be interpreted as a hint that
we do not suffer from massive ghost states.  However, a fully
conclusive investigation requires the access to the pole structure of
the graviton propagator, and hence a Wick rotation. Progress in the direction of real-time flows in general theories and gravity has been made e.g.\ in 
\cite{Floerchinger:2011sc,Kamikado:2013sia,Tripolt:2013jra,Pawlowski:2015mia,Strodthoff:2016pxx,Bonanno:2004sy,Manrique:2011jc,Biemans:2016rvp,Wetterich:2017ixo}.

\subsection{Higher order vertices and the background effective action}\label{sec:high+back}

The results in the last section immediately lead to the question about
the importance of the higher-order covariant tensor structures like
e.g.\ $f_{R^n}$ which have no overlap with the graviton $n$-point
functions computed in this work.  These are potentially relevant for
the flows of $G_5$ and $G_6$. These tensors have been dropped in the
current work, thus closing our vertex expansion. However, we may
utilise previous results obtained within the background field
approximation for estimating their importance: first we note that
$\R^2$ gives rise to a new relevant direction, as we will show in
\autoref{sec:UV-FP}. This has also been observed for the background
field
approximation~\cite{Lauscher:2002sq,Codello:2007bd,Machado:2007ea,Codello:2008vh,Falls:2013bv}.
There it has also been shown that the critical dimensions of the
$R^n$-terms approximately follow their canonical
counting~\cite{Falls:2013bv}. Furthermore, our results so far have
sustained the qualitative reliability of the background field
approximation for all but the most relevant couplings. Indeed, it is
the background field-dependence of the regulator that dominates the
deviation of the background approximation from the full analysis for
the low order vertices, and in particular the mass parameter $\mu$ of
the graviton. This field-dependence is less relevant for the higher
order terms. Thus, we may qualitatively trust the background field
approximation for higher curvature terms. This means that they are of
sub-leading importance and can be dropped accordingly.

Finally, the above findings together with those from the literature 
suggest that an Einstein-Hilbert action is generating a
diffeomorphism-invariant $\R^2$-term but not an
$\RicciTensor_{\mu\nu}^2$ term in the
diffeomorphism-invariant background effective action
$\Gamma_k[g]=\Gamma_k[g,\phi=0]$. Moreover, no higher derivative terms
are generated if a non-trivial wave function renormalisation
$Z_h(p^2)$ and graviton mass parameter $\mu=-2\lambda_2 $ are taken
into account. Note that this only applies for an expansion with
$p^2 < k^2$.  This is a very
interesting finding as it provides strong non-trivial support for the
semi-quantitative reliability of the background approximation in terms
of an expansion in $R$ for spectral values smaller than $k^2$ subject
to a resolution of the fluctuating graviton propagator: $\mu$ and
$Z_h$ have to be determined from the flows of the fluctuation fields
or in terms of the mNIs.

\subsection{Flow equations for the couplings} \label{sec:flows-couplings}
In this section we derive the flow equations for the
couplings from the projected $n$-point functions.

The flow equations for $\mass$ and
$\anomalousDimension{\graviton}(p^2)$ are extracted from the
transverse-traceless part of the flow of the graviton two-point
function.  We evaluate this two-point function at $p^2 = 0$ for
$\scaleDerivative{\mass}$, and bilocally at $- \mass k^2$ and $p^2$
for $\anomalousDimension{\graviton}(p^2)$. The algebraic equation for
$\anomalousDimension{\ghost}(p^2)$ can be obtained directly from the
transverse part of the flow of the ghost two-point function. These
equations are derived in the same fashion as in
\cite{Christiansen:2014raa,Meibohm:2015twa}. For details see also
App.~\ref{app:derivation-eq}. 

In the case of the couplings $\lambdan{n}$ and $\gn{n}(p^2)$, we project
onto the flow of the graviton \nptfct{n}s.  The flow equations for the
couplings $\lambdan{n}$ are always obtained at $p^2 = 0$, since
$\lambdan{n}$ describes the momentum-independent part of the graviton
\nptfct{n}s.

In the case of the couplings $\gn{n}(p^2)$ it is technically
challenging to resolve the full momentum dependence in the flow. Thus, we resort 
to a further approximation of the momentum-dependence. 
We have
checked that this approximation holds quantitatively. First we note
that typically FRG-flows are strongly peaked at $q\approx k$ due to
the factor $q^3$ from the loop integration and the decay for
momenta $q\gtrsim k$ due to $\partial_t R_k(q^2)$. This certainly holds
for all the flows considered here. From this we can infer that we extract the
leading contribution to the flow diagrams if we feed $\gn{n}(k^2)$
back into the diagrams.
In consequence we compute only the flow equations for $\gn{n}(k^2)$, as 
they form a closed system of equations within the given approximation. 

Conveniently, the momentum dependence of the flow for $\gn{3}(p^2)$ is
trivial, see \autoref{fig:mom_dep_flow} in
\autoref{sec:mom-dep-nptfct}.  Hence the approximation discussed above
is of quantitative nature, and we obtain precisely the same equation
as in \cite{Christiansen:2015rva}.

In contrast, the flow of the graviton four-point function exhibits a
$p^4$ contribution, implying a non-trivial $\gn{4}(p^2)$. Here our
approximation is necessary to simplify the computation significantly.
The flow equation for $\gn{4}(k^2)$ is obtained from a bilocal
momentum projection at $p^2=0$ and $p^2=k^2$, and furthermore uses an
approximation that relies on the fact that the coupling $\lambdan{4}$
remains small. We refer to this equation as a bilocal equation. It is
explicitly displayed in App.~\ref{app:derivation-eq}, see \eqref{eq:g4-k2-flow}.
Within our setup this equation gives
the best approximation of the vertex flows since it feeds back the
most important momentum information into the flow. This further
entails that the coupling $\gn{4}(k^2)$ includes information about the
invariants $R$ and $R^2$.

\subsection{Disentangling \texorpdfstring{$R$}{R} and
  \texorpdfstring{$R^2$}{R squared} tensor
  structures}\label{sec:R2-coupling}
In this section we present projection operators that disentangle
contributions from $R$ and ${R^2}$ tensor structures to the flows of
the couplings $\gn{n}(p^2)$. In the present setup this only allows us
to switch off the $R^2$ coupling and thus to check the importance
of the $R^2$ coupling.

For the disentanglement, we have to pay attention to two things: First
of all, a local momentum projection at $p^2=0$ is very sensitive to
small fluctuations and in consequence not very precise with regard to
the whole momentum range $0\leq p^2 \leq k^2$.  This was already
discussed in~\cite{Christiansen:2014raa,Christiansen:2015rva} and is
explicitly shown in App.~\ref{app:local-momentum-projection}.
Hence, we have to rely on non-local momentum projections.  Here the
highest polynomial power of $p^2$, as indicated in 
\eqref{eq:flow_polynomial}, dictates the simplest way of projecting on
the $p^2$-coefficient.  The graviton three-point function is at most
quadratic in the external momentum, and consequently it is enough to
use a bilocal projection at $p^2 = 0$ and $p^2 = k^2$. The resulting
equation is displayed in App.~\ref{app:derivation-eq}, see \eqref{eq:g3_flow}. 

The graviton
four-point function, on the other hand, has $p^4$ as its highest
momentum power, i.e.\ it is of the form
\begin{align} \label{eq:p4-polynom}
 f(p^2) &=  b_0 + b_1\, p^2 + b_2\, p^4 \,,
\end{align}
see also \eqref{eq:flow_polynomial}. Thus a bilocal momentum
projection would not extract the $p^2$ coefficient $b_1$ alone.
Instead, we use a trilocal momentum projection at $p^2=0$,
$p^2=k^2/2$, and $p^2=k^2$ in order to solve the above equation for
$b_1$.  I.e., we solve a system of linear equations and obtain
\begin{align} \label{eq:trilocal_mom_proj_p2}
  b_1 &= - 3 f(0) + 4 f(k^2/2) - f(k^2) \,.
\end{align}
The resulting flow equation is again presented in
App.~\ref{app:derivation-eq}, see \eqref{eq:g4_flow}.

For even higher order momentum contributions we would have to use even
more points of evaluation.  These momentum projections together with
the observation of \eqref{eq:flow_polynomial} guarantee that
we project precisely on the $p^2$ coefficient in the whole momentum
range $0\leq p^2 \leq k^2$.

A natural upgrade of the current approximations amounts to the
introduction of a second tensor structure that is orthogonal to the
Einstein-Hilbert one in terms of these projections. Within our
uniformity assumption this is considered to be sub-leading, and the
momentum-dependence of $g_4(p^2)$ takes care of the contribution
of the $R^2$ tensor structure $f^{(4)}_{R^2}$. While the orthogonal
projection on the respective flow is simple, its back-feeding demands
a two tensor structure approximation of the three- and four-graviton
vertex in the flow, the implementation of which is deferred to future
work.

Here, we only perform a further check of the relevance of the
$R^2$ tensor structure. This sustains the fact that the inclusion of
the four-graviton vertex with its contribution of the $R^2$ tensor
structure leads to an additional UV-relevant direction. To that end we
generalise our ansatz for the graviton four-point function such that
we can extract a flow equation for both the Einstein-Hilbert tensor
structure as well as for the $\R^2$ tensor structure. As already
mentioned above, we cannot feed the generated coupling back into the
flows, since they are given by vertices with Einstein-Hilbert tensor
structures. Instead we compute the fixed point value that arises only
from the Einstein-Hilbert tensor structures.

As the corresponding ansatz for the transverse-traceless graviton
four-point function we choose
\begin{align}
  \actionQu_k^{(4)}(p^2) &= \wfr[2]{h}{p^2} \Gn{4} \left(
    \constantGnLambda{4} \Lambdan{4} + \constantGnp{4} p^2 +
    \constantGnRsquared{4} \RsquaredCouplingndim{4} p^4 \right) \,,
\end{align}
which is precisely the vertex that emerges from the sum of 
Einstein-Hilbert tensor structure and $\R^2$ tensor structure. The
related generating diffeomorphism-invariant action for this
four-graviton vertex is
\begin{align}\label{eq:EH_action_R2}
  \actionCl &= \actionEH + \frac{1}{16 \pi \newtonG} \int \dn{4}{x}
  \sqrt{\det g} \, \RsquaredCouplingdim\, \R^2 \,,
\end{align}
where $\actionEH$ is defined as in \eqref{eq:EH_action}. The
flow of $\Omega_4$ is then obtained by the trilocal momentum
projection described below \eqref{eq:p4-polynom}.  For $b_2$ we obtain
\begin{align} \label{eq:trilocal_mom_proj_p4}
  b_2 &= 2 f(0) -4 f(k^2/2) +2 f(k^2) \,.
\end{align}
The explicit form of the resulting flow equation for the dimensionless
coupling $\RsquaredCouplingn{4} \definition \RsquaredCouplingndim{4}
k^2$ is given in App.~\ref{app:derivation-eq}, see \eqref{eq:o4_flow}.  Note that in the
present approximation, the flows do not depend on the coupling
$\RsquaredCouplingn{4}$ since it does not feed back into the vertices.

\subsection{Computational details}
The computations of correlation functions described in this section
involve contractions of very large tensor structures.  To give a rough
estimate: the classical Einstein-Hilbert three-point vertex alone
consists of around 200 terms, and the classical graviton propagator of
7 terms.  For the box diagram of the flow of the graviton four-point
function, displayed in \autoref{fig:flow_n_point}, this results in a
total number of approximately $200^4\cdot 7^4 \approx 4\cdot10^{12}$
terms, if no intermediate simplifications are applied.

These contractions are computed with the help of the symbolic
manipulation systems {\small
  \emph{FORM}}~\cite{Vermaseren:2000nd,Kuipers:2012rf} and
\emph{Mathematica}~\cite{mathematica}.  For individual tasks, we
employ specialised and in part self-developed Mathematica packages.
In particular, we use \emph{VertEXpand}~\cite{vertexpand} and
\emph{xPert}~\cite{xPert} for the generation of vertex functions,
\mbox{\emph{DoFun}}~\cite{DoFun} to obtain symbolic flow equations,
and the
\emph{FormTracer}~\cite{Cyrol:2016zqb,Mitter:2014wpa,Cyrol:2016tym} to
create optimised {\small \emph{FORM}} scripts to trace diagrams.
Furthermore, we make use of the self-developed framework
{\small\emph{TARDIS}}~\cite{TARDIS} facilitating an automated and
seamless usage of the aforementioned tools.

\section{Asymptotic safety}\label{sec:AS}
In this section, we discuss the UV fixed point structure of our
system.  We first present our best result, which includes the tensor
structures as presented in \autoref{sec:invariants} and in particular 
in \eqref{eq:Rf} and \eqref{eq:g3-tensors}. The underlying UV-relevant diffeomorphism
invariants turn out to be $\Lambda$, $R$, and $R^2$.  The $R^2$ coupling is included
via the momentum dependence of the gravitational coupling
$\gn{4}(p^2)$, see \autoref{sec:flows-couplings}.  As a main result we
find an attractive UV fixed point with three attractive directions.
The third attractive direction is related to the inclusion of the
$R^2$ coupling.

We further analyse the stability of this UV fixed point with respect to
the identification of the higher couplings.  We also analyse the
previous truncation \cite{Christiansen:2015rva} and compare the
stability of both truncations.  Here we find that the improvement of
the truncation increases the stability of the system.  In particular,
we find a rather large area in the theory space of higher couplings
where the UV fixed point exists with three attractive directions
throughout.

Lastly, we discuss the importance of the $R^2$ coupling.  In
\autoref{sec:R2-coupling} we have constructed projection operators
that disentangle the contributions from $R$ and $R^2$ tensor
structures. This allows us to switch off the $R^2$ coupling and
compare the stability of the reduced system to that of the full
system.  We find that the reduced system is significantly less stable,
and that the area in the theory space of higher couplings where the
fixed point exists is rather small.  This highlights the importance of
the $R^2$ coupling.

\begin{figure*}[tbp]
  \centering
  \includegraphics[width=0.47\textwidth]{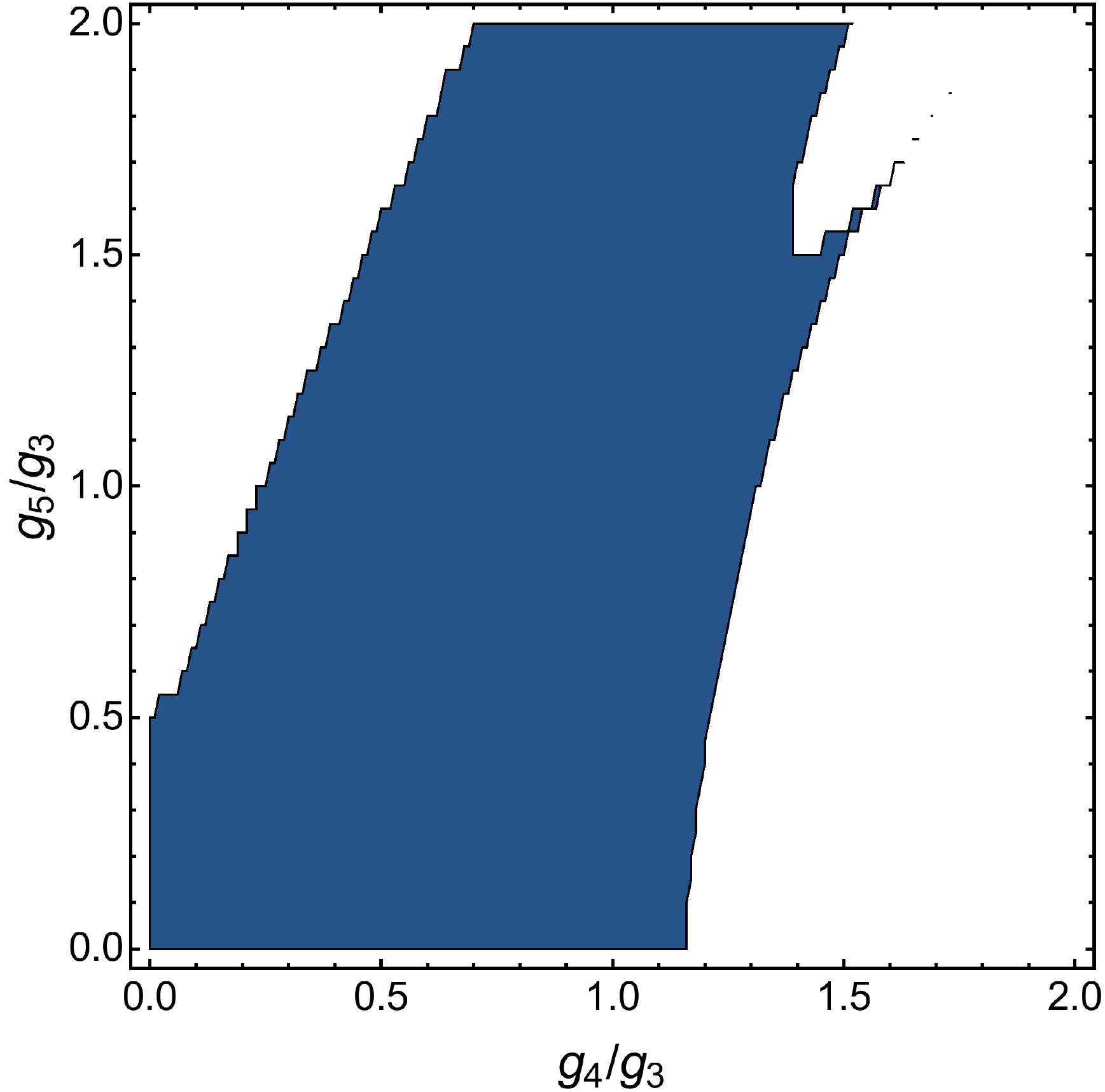}
  \hspace{.5cm}
  \includegraphics[width=0.47\textwidth]{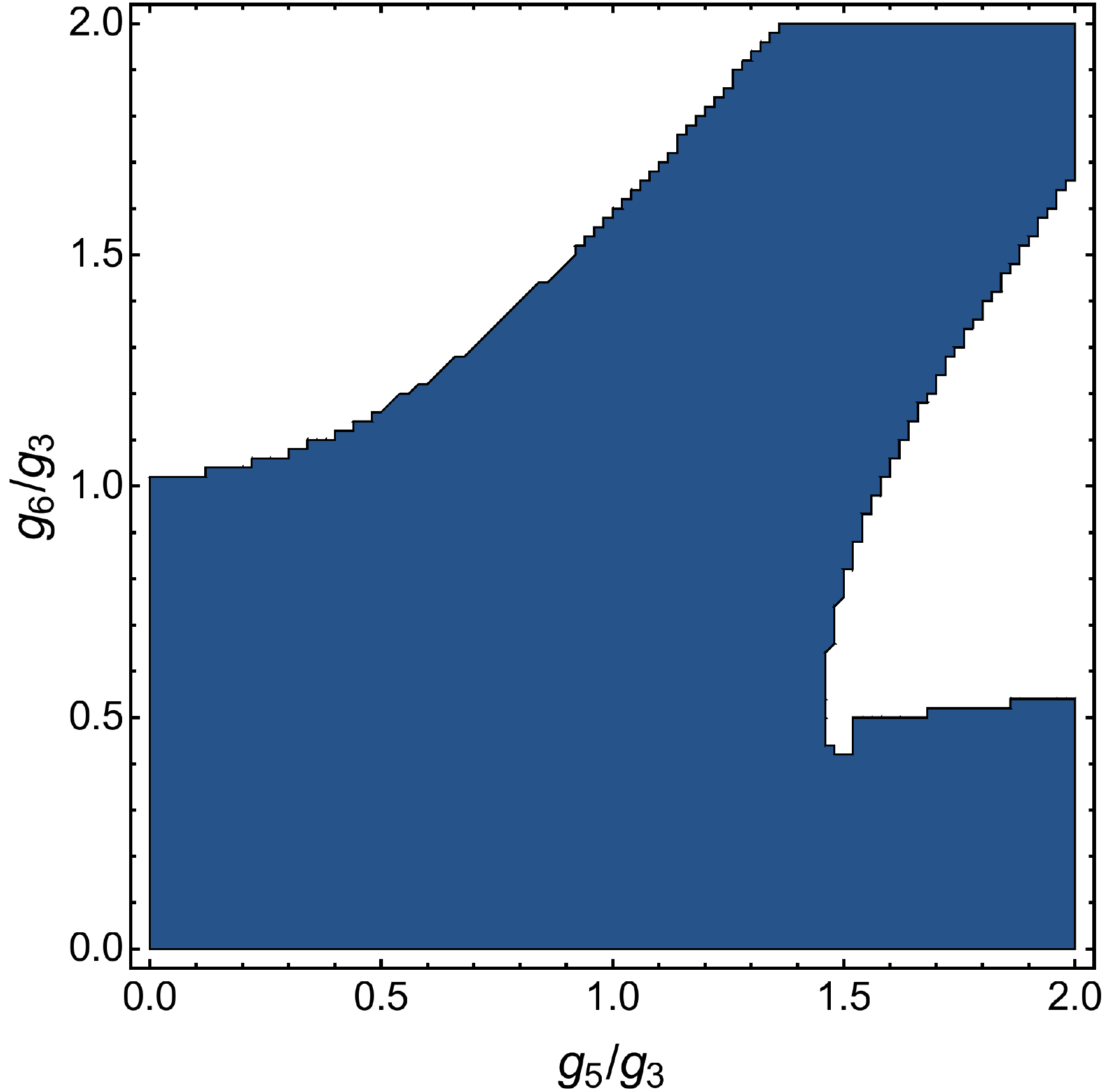}
  \caption[gn grid]
  {Plots of the existence of an attractive non-trivial UV fixed point
    (blue) dependent on the higher couplings.  Left: The system
    $(\mu,\lambdan{3},\gn{3})$ \cite{Christiansen:2015rva} dependent
    on the higher couplings $\gn{4}$ and $\gn{5}$.  Right: The system
    $(\mu,\lambdan{3},\lambdan{4},\gn{3},\gn{4})$ dependent on the
    higher couplings $\gn{5}$ and $\gn{6}$.  The higher couplings
    $\lambdan{n>n_{\mathrm{max}}}$ are always identified with
    $\lambdan{3}$.  The blue area marks the region where an attractive
    UV fixed point was found.  At the border of this area the fixed
    point either vanishes into the complex plane or loses its
    attractiveness.  In both systems the area where the fixed point
    exists is rather large and contains the identification $\gn{n>n_{\mathrm{max}}}=\gn{3}$. 
    Conveniently, the area increases for the better
    truncation, indicating that the system becomes more stable with an
    improvement of the truncation.  The number of attractive
    directions is uniformly two in the left panel and three in the
    right panel.}
  \label{fig:fp-window}
\end{figure*}

\subsection{UV fixed point}\label{sec:UV-FP}
In this section we display the UV fixed
point structure of our full system.  This means that we feed back the
generated $\R^2$ coupling via the momentum dependence of the
gravitational coupling $\gn{4}(p^2)$, as discussed
in~\autoref{sec:flows-couplings}. Fixed
points are by definition points where the flows of the dimensionless
couplings vanish. In consequence, we look for the roots of
the equations \eqref{eq:mu_dot}, \eqref{eq:lamn_dot},
\eqref{eq:g3_flow}, and \eqref{eq:g4-k2-flow}.  We use the identification scheme
$\gn{6}=\gn{5}=\gn{4}$ and $\lambdan{6}=\lambdan{5}=\lambdan{3}$.  We
find a UV fixed point at the values
\begin{align} \label{eq:UV-FP} \left( \mu^*, \lambdan{3}^*,
    \lambdan{4}^*, \gn{3}^*, \gn{4}^* \right) &= \left( -0.45,\, 0.12,\,
    0.028,\, 0.83,\, 0.57 \right) \,.
\end{align}
The fixed point values are similar to those of the previous truncation
\cite{Christiansen:2015rva}.  The biggest change concerns the graviton
mass parameter, which is now less negative and thus further away from
its pole.  Moreover, it is remarkable that the new couplings
$\lambdan{4}$ and $\gn{4}$ are close to their lower counterparts
$\lambdan{3}$ and $\gn{3}$, but not at precisely the same values.
Since we use the difference between these couplings to parameterise
the breaking of diffeomorphism invariance, this is more or less what
we expected.  This issue is further discussed in the next section.

We do not have access to the full stability matrix of the UV fixed
point due to the unknown flow equations of the higher couplings.  For
this reason, we discuss two different approximations of the stability
matrix.  The main difference between these two approximations concerns
the order of taking the derivatives and identifying the higher
couplings, which is explained in more detail in
App.~\ref{app:stability-matrix}. We argue that in a well
converged approximation scheme the most relevant critical exponents
should not depend on the approximation of the stability matrix.  Thus,
we can use the two different approximations to judge the quality of
the current level of truncation.
In this work, we define the critical exponents as the eigenvalues of the stability matrix without a minus sign.
We call the critical exponents of
the first approximation $\bar \criticalExponent_i$, and the ones of the
second approximation $\tilde \criticalExponent_i$.  The critical
exponents using the first approximation are given by
\begin{align}
   \bar  \criticalExponent_i &= (-4.7,\; -2.0 \pm 3.1 \imaginaryi,\; 2.9,\; 8.0 ) \,, \label{eq:crit_exp_w_bf}
\end{align}
while the critical exponents using the second approximation are
\begin{align}
    \tilde \criticalExponent_i &= (-5.0,\; -0.37 \pm 2.4 \imaginaryi,\; 5.6,\; 7.9) \,.
\end{align}
Hence this fixed point has three attractive directions in both
approximations of the stability matrix.  The third attractive
direction compared to the system of the graviton three-point function
\cite{Christiansen:2015rva} is related to the fact that the
graviton four-point function has an overlap with $R^2$, which we feed
back via the momentum dependence of the gravitational coupling
$\gn{4}(p^2)$.  The $R^2$ coupling has also been relevant in earlier
computations with the background field approximation
\cite{Lauscher:2002sq,Codello:2007bd,Machado:2007ea,Codello:2008vh,Falls:2013bv}.
In addition, note that the most attractive eigenvalue is almost identical in
both approximations of the stability matrix.  This is a positive sign
towards convergence since it is expected that the lowest eigenvalue is
the first that converges, c.f.\ App.~\ref{app:stability-matrix}.

Furthermore, the anomalous dimensions at the UV fixed point read
\begin{align} \nonumber
 (\anomalousDimension{\graviton}^*(0), \anomalousDimension{\graviton}^*(k^2)) &= (0.56  ,\; 0.079 )\,, \\[1ex]
 (\anomalousDimension{\ghost}^*(0), \anomalousDimension{\ghost}^*(k^2)) &= ( -1.28 ,\; -1.53 )\,,
\end{align}
where we have chosen to display the anomalous dimensions at the
momenta that feed back into the flow.  All anomalous dimensions stay
well below the reliability bound $\eta_{\phi_i}(p^2)<2$, as introduced
in~\cite{Meibohm:2015twa}.

\subsection{Stability} \label{sec:stability}
In the following we investigate the UV fixed point from the previous
section by varying the identification of the higher couplings.
Again we look for the roots of the equations
\eqref{eq:mu_dot}, \eqref{eq:lamn_dot}, \eqref{eq:g3_flow}, and
\eqref{eq:g4-k2-flow}.  These equations however still depend on the
higher couplings $\gn{5}$, $\gn{6}$, $\lambdan{5}$, and $\lambdan{6}$.
We have to identify these couplings with the lower ones or set them to
constants in order to close the flow equations.

It is a natural choice to simply set these higher couplings equal to
lower ones, e.g.\ $\gn{6}=\gn{5}=\gn{3}$ and
$\lambdan{6}=\lambdan{5}=\lambdan{3}$, as done in the previous
section.  The couplings would fulfil this relation exactly in a fully
diffeomorphism invariant setup.  However, such a diffeomorphism
invariant setup is not at hand.  In fact, we can parameterise the
breaking of diffeomorphism invariance via these couplings, e.g.\ by
writing $\gn{n} = \gn{3} + \Delta_{g_n}$.  Here we have designated
$\gn{3}$ as a reference coupling since it is the lowest genuine
gravitational coupling.  For this reason, it is also the most
converged gravitational coupling within this vertex expansion, thus
justifying this choice.  In general we expect $\Delta_{g_n}$ to be
small and in consequence we vary the identification of the higher
couplings only in this part of the theory space of higher couplings.
The quantity $\Delta_{g_4}$ is indeed small at the UV fixed point
presented in the last section, see \eqref{eq:UV-FP}.
More precisely, it takes the value $|\Delta_{g_4}/\gn{3}| \approx 0.3$ 
at this UV fixed point.

In this analysis we choose to identify 
\begin{align}
  \gn{5} &= \alpha_1\, \gn{3} \,, &
  \gn{6} &= \alpha_2\, \gn{3} \,,
\end{align}
and $\lambdan{6}=\lambdan{5}=\lambdan{3}$ for simplicity, and investigate the existence of the UV fixed point as
a function of the parameters $\alpha_1$ and $\alpha_2$.  In
\autoref{fig:fp-window} the area where an attractive UV fixed point
exists is displayed in blue.  In the left panel, this is done for the
previous truncation
$(\mu,\lambdan{3},\gn{3})$~\cite{Christiansen:2015rva}, and in the
right panel for the current truncation
$(\mu,\lambdan{3},\lambdan{4},\gn{3},\gn{4})$. At the border of the
blue area the UV fixed point either vanishes into the complex plane or
loses its attractiveness.  Remarkably, both areas are rather large,
suggesting that the existence of the UV fixed point is quite stable.
Even more conveniently, the area increases with the improved
truncation, suggesting that the system is heading towards a converging
limit.  Note that the number of attractive directions of the UV fixed
point is constant throughout the blue areas, namely two in the left
panel and three in the right panel.

We further analyse the fixed point values that occur within the blue
area in the right panel of \autoref{fig:fp-window}.  Interestingly,
the fixed point values are rather stable throughout the whole area
where the UV fixed point exists.  More precisely, they stay within the
following intervals:
\begin{align}
  \mu^* &\in [-0.72,\,-0.19] \, \nonumber\\[1ex]
  \lambdan{3}^* &\in [-0.018,\,0.29] \,\nonumber\\[1ex]
  \lambdan{4}^* &\in [-1.2,\,0.12] \,\nonumber\\[1ex]
  \gn{3}^* &\in [0.22,\,1.4] \, \nonumber\\[1ex]
  \gn{4}^* &\in [0.11,\,0.97] \,. \nonumber
\end{align}
Hence, in particular the fixed point value of $\lambdan{3}$ is already
confined to a very small interval, and also a very small number.  The
latter is important since some of our approximations rely on the fact
that the $\lambdan{n}$ are small, see \autoref{sec:flows-couplings}.
The fact that $\lambdan{4}^*$ is varying more strongly than
$\lambdan{3}^*$ is not surprising since we expect $\lambdan{3}$ to be
better converged, being a lower coupling.  The fixed point values of
$\gn{3}$ and $\gn{4}$ seem to try to compensate the change induced by
the identification.  Thus, $\gn{3}^*$ and $\gn{4}^*$ become larger
towards the identification $\gn{6}=\gn{5}=0$ and smaller towards
$\gn{6}=\gn{5}=2\gn{3}$.  The shape of the area in the left panel in
particular suggests the relation $\gn{4}^*<\gn{3}^*$, which is
fulfilled by the improved truncation almost throughout the whole area
where the fixed point exists.  This is indeed a non-trivial prediction
that has been fulfilled by our approximation scheme.

A further study of the dependence of the UV fixed point properties on
the choice of identification is given in
App.~\ref{app:identification-scheme}.

\begin{figure}[!tb]
  \centering
  \includegraphics[width=.95\linewidth]{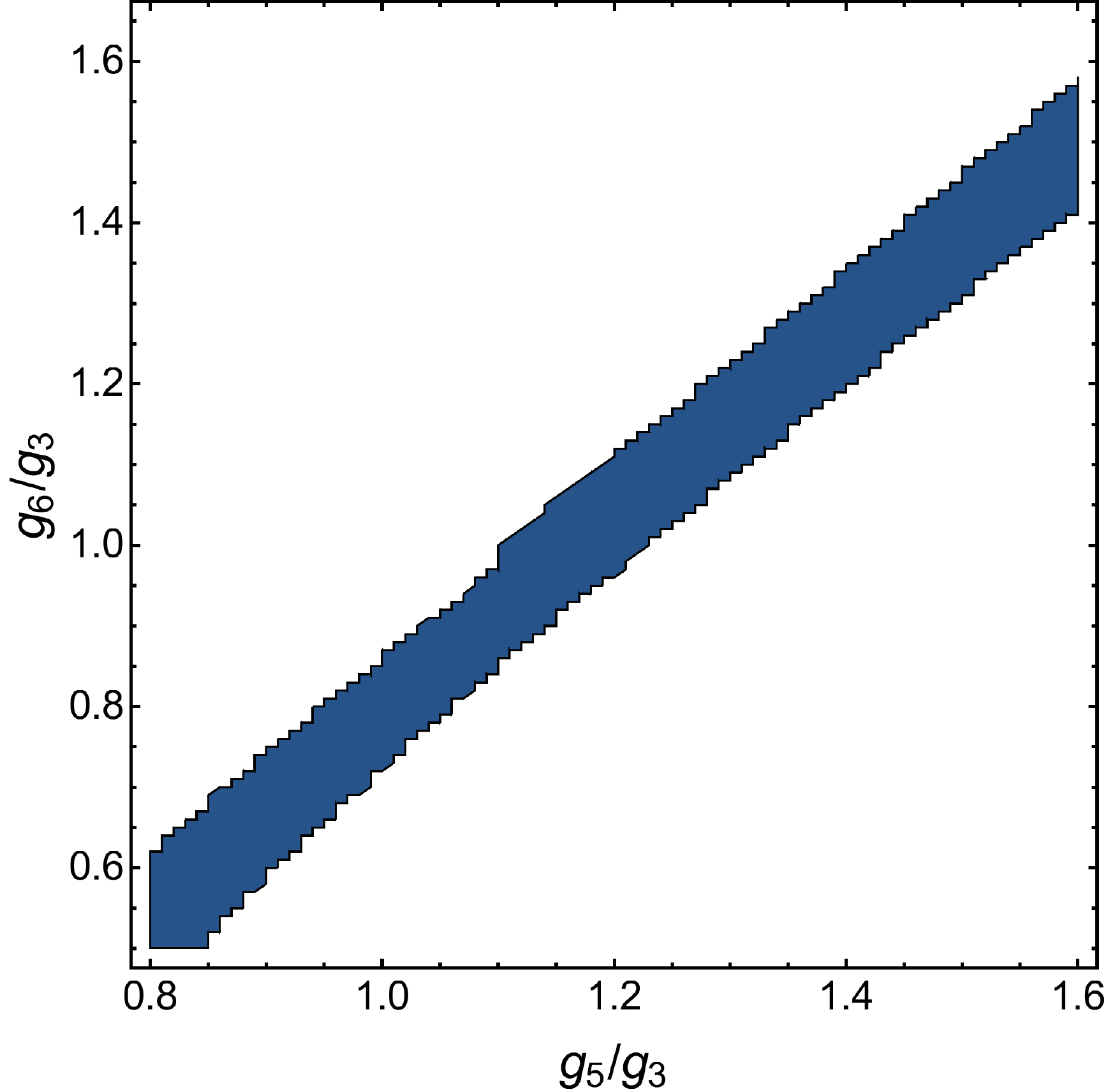}
  \caption[gn grid trilocal]{Plot of the existence of an attractive
    non-trivial UV fixed point (blue) dependent on the higher
    couplings $\gn{5}$ and $\gn{6}$.  Here, the trilocal equation for
    the gravitational coupling $\gn{4}$ was used, which 
    allows us to switch off the
    $\R^2$ coupling. We found two different
    fixed points with rather similar fixed point values. Each fixed point has 
    its own area of existence in the theory space of the higher couplings. 
    The blue area marks the unified area of both fixed points. 
    Nevertheless, the area is
    significantly smaller than the areas displayed
    in \autoref{fig:fp-window}.  This reflects the importance of
    the $\R^2$ coupling.}
  \label{fig:fp-window-tl}
\end{figure}

\subsection{Importance of the \texorpdfstring{$R^2$}{R squared} tensor structure}\label{sec:relevance-R2}
In the previous subsection we have fed back the $R^2$ contributions to
the flow via the momentum-dependent gravitational coupling
$\gn{4}(p^2)$. In order to check the quality of our approximation and
to investigate the influence of the $\R^2$ tensor structure on the
fixed point structure of the system, we switch off the $R^2$ contribution in this section.
We do the latter by projecting onto the $p^2$ part of the
flow via a trilocal momentum projection scheme,
cf.~\autoref{sec:mom-dep-nptfct} and \autoref{sec:R2-coupling}.
This is
both an examination of the influence of $\R^2$ on the results
presented in the previous subsections, as well as a proof of concept
for disentangling the tensor structures of different invariants.
Our analysis in this subsection suggests that
leaving out the contribution of $\R^2$ leads to significantly less stable results.

\begin{figure*}[htb!]
  \centering
  \includegraphics[width = \textwidth]{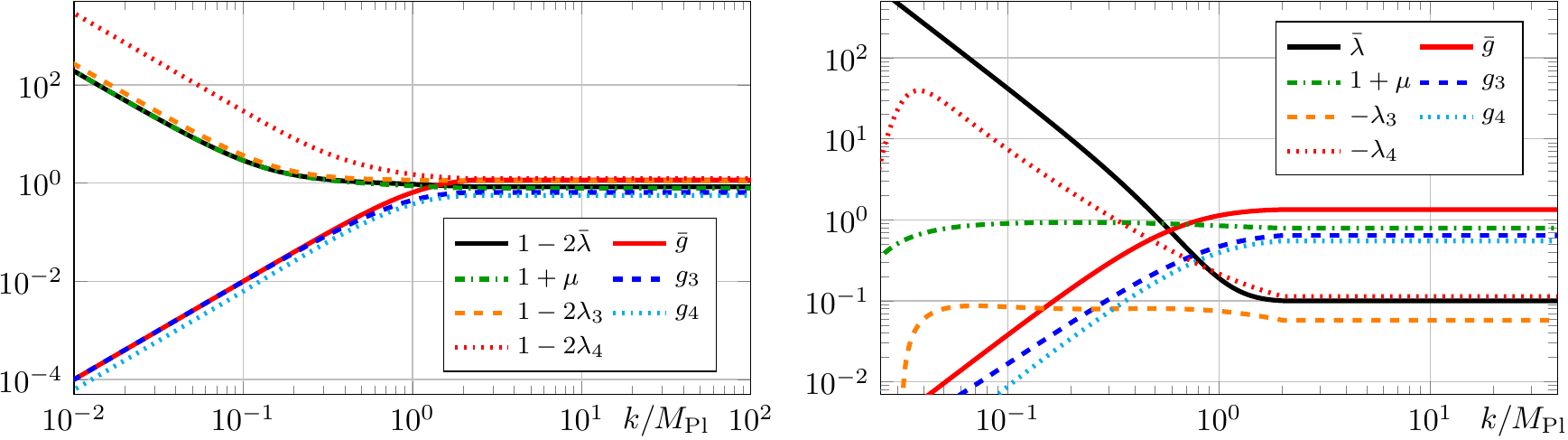}
  \caption[IR trajectories]{ Examples of UV finite trajectories from
    the UV fixed point \eqref{eq:UV-FP} towards the IR. In the left
    panel all couplings scale classically below the Planck scale and
    reach their UV fixed point values shortly above the Planck scale.
    In the right panel some couplings show non-classical behaviour
    even below the Planck scale, which is triggered by the graviton
    mass parameter $\mass$ flowing towards the pole of the graviton
    propagator at $\mass = -1$. However, in this case the numerics
    break down at $k \approx 0.02 \massPlanck$ due to competing orders
    of the factor $(1 + \mu)$ close to the singularity at $\mass =
    -1$. The trajectories in both panels correspond to theories that behave like
    classical general relativity in the IR.  Note that some couplings
    are plotted shifted or with a minus sign in order to keep them
    positive over the whole range.}
  \label{fig:trajectory_IR_FP}
\end{figure*}

In \autoref{fig:fp-window-tl} we display the result for the same
analysis as in the previous section, but with the trilocal equation \eqref{eq:g4_flow} for
$\gn{4}$ instead.
We find two fixed points with rather similar fixed point values.
However, we are only interested in identifying the area in the theory space of the higher couplings where at least one UV fixed point exits.
Thus, we unify both areas and obtain the blue area displayed in \autoref{fig:fp-window-tl}.
This area forms a rather narrow band whose total area is significantly smaller than for the momentum dependent gravitational
coupling $\gn{4}(p^2)$, c.f.~\autoref{fig:fp-window}.
The identification $\gn{6}=\gn{5}=\gn{3}$ also does not lie within these regions, but just outside of them.
Since we switched off the $\R^2$ contribution, a less stable fixed point structure was to be expected,
and consequently these results highlight the importance of the $\R^2$ coupling.

\begin{table*}
  \centering
  \caption[UV fixed points for different systems of beta functions]{
  Properties of the non-trivial UV fixed point for different orders of the vertex expansion scheme, 
  computed for momentum dependent anomalous dimensions $\eta_{\phi_i}(p^2)$ and bilocally projected Newton's couplings $\gn{n}(k^2)$. 
  The critical exponents $\bar \criticalExponent_i$ and $\tilde \criticalExponent_i$ 
  stem from two different approximation of the stability matrix as discussed in App.~\ref{app:stability-matrix}.
  The fixed points are computed with the identifications $\gn{6}=\gn{5}=\gn{\text{max}}$ and $\lambdan{6}=\lambdan{5}=\lambdan{3}$.
  We observe that the fixed point values are only varying mildly between the different orders of the vertex expansion.
  Notably, if we compare the critical exponents of the two approximations of the stability matrix,
  we observe that the difference becomes smaller with an increasing order of the vertex expansion.
  This is precisely what one would expect of a systematic approximation scheme that is approaching a converging limit.
  }
  \label{tab:UV_fixed_points_truncations_overview}
  \vspace{5pt}
  \setlength{\tabcolsep}{3.5pt}
  \begin{tabular}{ l @{\hskip 25pt} c  c  c  c  c  @{\hskip 25pt} c  c  c  c  c  @{\hskip 25pt} c  c  c  c  c }
    \toprule
      System &  $\mu^*$ & $\lambdan{3}^*$ & $\lambdan{4}^*$ & $\gn{3}^*$ & $\gn{4}^*$ & \multicolumn{5}{c}{$\bar \criticalExponent_i$}  &  \multicolumn{5}{c}{$\tilde \criticalExponent_i$}  \\
    \midrule
      $\mu$, $\gn{3}$, $\lambdan{3}$ & $-0.57$ & $0.095$ & & $0.62$ & & \multicolumn{2}{c}{$ -1.3 \pm 4.1\imaginaryi$} & $12$ & & & $-7.3$ & $3.5$ & $7.4$ & &  \\ 

      $\mu$, $\gn{3}$, $\lambdan{3}$, $\gn{4}$ & $-0.53$ & $0.086$ & & $0.74$ & $0.67$ & \multicolumn{2}{c}{$-2.1 \pm 3.8 \imaginaryi$} & $3.6$ & $11$ &  &  \multicolumn{2}{c}{$-0.75 \pm 1.5 \imaginaryi$} & \multicolumn{2}{c}{$7.8 \pm 3.5\imaginaryi$} & \\ 

      $\mu$, $\gn{3}$, $\lambdan{3}$, \phantom{$\gn{4}$,} $\lambdan{4}$ & $-0.58$ & $0.17$ & $0.032$ & $0.48$ & & $-4.1$ & \multicolumn{2}{c}{$-0.35 \pm 2.6 \imaginaryi$} & $8.3$ &  & $-6.2$ & $-1.8$ & $3.4$ & $8.8$ & \\

      $\mu$, $\gn{3}$, $\lambdan{3}$, $\gn{4}$, $\lambdan{4}$ & $-0.45$ & $0.12$ & $0.028$ & $0.83$ & $0.57$ & $-4.7$ & \multicolumn{2}{c}{$-2.0 \pm 3.1 \imaginaryi$} & $2.9$ & $8.0$ & $-5.0$ & \multicolumn{2}{c}{$-0.37 \pm 2.4 \imaginaryi$} &  $5.6$ & $7.9$ \\
    \bottomrule
  \end{tabular}
\end{table*}

\section{IR behaviour}\label{sec:IR-behaviour}
In this section, we discuss the IR behaviour of the present theory of
quantum gravity.  We only consider trajectories that lie within the UV
critical hypersurface, i.e.\ trajectories that are UV finite, and which end
at the UV fixed point presented in \eqref{eq:UV-FP} for $k\to
\infty$. In this section we use the analytic flow
equations given in App.~\ref{app:analytic-eq} for simplicity, and set the
anomalous dimensions to zero, i.e.\ $\anomalousDimension{\phi}=0$.
This approximation gives qualitatively similar results, as discussed
in App.~\ref{app:local-momentum-projection}.

In the IR, it is particularly interesting to examine the background
couplings $\backgroundg$ and $\backgroundLambda$.  In the limit $k\to
0$ the regulator vanishes by construction and the diffeomorphism
invariance of the background couplings is restored.  Hence they become
observables of the theory.  The flow equations for the background
couplings are displayed in App.~\ref{app:background_couplings}.

In general we look for trajectories that correspond to classical
general relativity in the IR.  This implies that the quantum
contributions to the background couplings vanish and in consequence
that they scale classically according to their mass dimension.  The
classical scaling is described by
\begin{align} \label{eq:class-scaling} \backgroundg,\, \gn{3},\,
  \gn{4}\, \sim k^2, \quad\quad \backgroundLambda,\, \mass,\,
  \lambdan{3},\, \lambdan{4}\, \sim k^{-2} \,.
\end{align}
We use the classical scaling in the flow from the UV fixed point to
the IR in order to set the scale $k$ in units of the Planck mass
$\massPlanck$.  We need to find a large enough regime where
$\backgroundg \sim k^{-2}$.  This entails that Newton's coupling
is a constant in this regime and sets the scale $k$ via $\newtonG =
\massPlanck^{-2} = \backgroundg k^{-2}$.

In \autoref{fig:trajectory_IR_FP}, two exemplary trajectories are
displayed.  In the left panel all couplings scale classically below
the Planck scale and reach their UV fixed point values shortly above
the Planck scale.  All quantum contributions are suppressed simply by
the fact that $\mu\to\infty$.  In the right panel on the other hand
some couplings exhibit a non-classical behaviour even below the Planck
scale, which is triggered by the graviton mass parameter $\mass$
flowing towards the pole of the graviton propagator at $\mass = -1$.
This entails that the dimensionful graviton mass parameter
$\Mass^2=\mu k^2$ is vanishing in the IR.  This IR behaviour is
analogous to the one observed in \cite{Christiansen:2014raa}, and
recently also \cite{Biemans:2016rvp}.  Remarkably, not only $\mu$ is
behaving non-classically but also $\lambdan{3}$, even though it is not
restricted by any pole.  However, in this scenario the numerics break
down at $k \approx 0.02 \massPlanck$ due to competing orders of the
factor $(1 + \mu)$ close to the singularity at $\mass = -1$.

In the left panel we have tuned the background couplings
$\backgroundg$ and $\backgroundLambda$ so that they are equal to the
lowest corresponding fluctuation coupling in the IR,
i.e.\ $\backgroundg=\gn{3}$ and
$\backgroundLambda=\lambdan{2}=-\mass/2$ for $k\ll \massPlanck$.  This
is equivalent to solving a trivial version of the Nielsen identities
(NIs).  Since all quantum contributions are suppressed by the graviton
mass parameter going to infinity in the IR, $\mass\to\infty$, the NI
in \eqref{eq:NI} reduces to
\begin{align}
  \0{\delta \Gamma[\bar g,h]}{\delta \bar g} =\0{\delta \Gamma[\bar g,
    h]}{\delta h} \quad \quad \text{for} \quad \mu\to\infty \quad \&
  \quad k\to0 \,. \notag
\end{align}
In consequence, we should see that all couplings coincide in this
limit, $\backgroundg=\gn{n}$ and $\backgroundLambda=\lambdan{n}$.
This is not the case in the left panel of
\autoref{fig:trajectory_IR_FP} since we have only fine-tuned the
background couplings, and thus we have two further degrees of freedom
that could be used for fine-tuning, stemming from the three
dimensional UV critical hypersurface.  We postpone this fine-tuning
problem to future work.

In summary, we find different types of trajectories that correspond to
classical general relativity in the IR.  The main difference lies in
the behaviour of the graviton mass parameter $\mass$, which flows to
infinity in one case and to minus one in the other case.  Both
scenarios are equivalent to general relativity in the end, in
particular since only the background couplings become observables in
the limit $k\to 0$.

\section{Towards apparent convergence}\label{sec:convergence}
In this section we discuss and summarise the findings of this work
concerning apparent convergence.  On the one hand, the order of our
vertex expansion is not yet high enough to fully judge whether the
system approaches a converging limit.  Nevertheless, we have collected
several promising first hints that we want to present in the
following.

In this work we have introduced two different approximations to the
stability matrix, as presented in App.~\ref{app:stability-matrix}.
We have argued that in a well
converged approximation scheme the most relevant critical exponents
should not depend on the approximation of the stability matrix.
In \autoref{tab:UV_fixed_points_truncations_overview} we display the UV
fixed point properties for different orders of the vertex expansion.
The first system is without the graviton four-point function and
exactly the same as in \cite{Christiansen:2015rva}. Then we look at
systems where we add either only an equation for $\gn{4}(k^2)$
(c.f.~\eqref{eq:g4-k2-flow}), or only an equation for $\lambdan{4}$
(c.f.~\eqref{eq:lamn_dot}). Lastly, we display our best truncation
including all couplings up to the graviton four-point function, see \autoref{sec:UV-FP}.
We observe that the fixed point values of
the couplings vary only mildly with an improving truncation, although there
is no clear pattern to those variations.
The most important piece of
information is the difference between the critical exponents from the
two different approximations of the stability matrix.  While the
difference is rather large in the truncation of the graviton
three-point function, it is becoming smaller with each improvement of
the truncation.  At the level of the graviton four-point function, the
critical exponents show only a small difference. This is precisely
what we expect, and thus we interpret this as a sign that the system
is approaching a converging limit.

Another important piece of information comes from the stability of the
UV fixed point under different closures of the flow equation. In a
well converged expansion scheme, the properties of the UV fixed point
should be completely insensitive to the details of the closure of the
flow equation. We have performed this analysis in
\autoref{sec:stability}. We observed that the area in which the UV
fixed point exists in the theory space of higher couplings is indeed
increasing with the improvement of the truncation. Furthermore, we
saw that the UV fixed point values are confined to small intervals.
We again interpret this as a sign that the system is approaching a
converging limit.

In summary, we have already seen several signatures of apparent
convergence although we are only at the level of the graviton
four-point function within the present systematic expansion scheme.
This suggests that we are on a promising path and that the present
setup will eventually lead to a converging limit.

\section{Summary}\label{sec:summary}
We have investigated quantum gravity with a vertex
expansion and included propagator and vertex flows up to the graviton
four-point function.  The setup properly disentangles background and
fluctuation fields and, for the first time, allows to compare two
genuine Newton's couplings stemming from different vertex flows.
Moreover, with the current truncation we have closed the flow of the
graviton propagator: all vertices and propagators involved are computed 
from their own flows. 

As a first non-trivial result we have observed that the
vertex flows of the graviton three-point and four-point functions, 
in the sense of \eqref{eq:flow_polynomial}, are
well described by a polynomial in $p^2$ within the whole momentum
range $0\leq p^2 \leq k^2$. The projection used for the flows 
takes into account the $\R$, $\R^2$ and
$\RicciTensor_{\mu\nu}^2$ tensor structures as well
as higher order invariants with covariant momentum
dependencies. Importantly, it is orthogonal to the $\R^2$ tensor
structure for the graviton three-point function, but includes it for the
graviton four-point function. We have shown that the highest momentum power contributing to the
graviton three-point function is $p^2$.
Therefore, $\RicciTensor_{\mu\nu}^2$ and higher derivative terms do not contribute to the graviton three-point function.
Thus, in particular $\RicciTensor_{\mu\nu}^2$ is excluded as a UV-relevant direction.
On the other hand, the flow of the graviton four-point function shows $p^4$ as its highest momentum power.
Together with the three-point function result
we infer that $\R^2$ is UV-relevant and contributes to the graviton four-point function.
This is a very interesting and highly non-trivial result.

At the moment, we cannot make final statements about higher $\R^n$
terms directly from our analysis. Nonetheless, predictions can be made
with a combination of the results presented here and previous ones obtained within
the background field approximation as well as the vertex expansion: Firstly, 
our work sustains the qualitative reliability of background field or
mixed approximations for all but the most relevant couplings. We have
seen that the range of allowed Newton's couplings stemming from
$n$-graviton vertices is growing with the level of the
approximation. Moreover, in \cite{Eichhorn:2016esv,Meibohm:2015twa} it has been shown
that already the substitution of the most relevant operator, the mass
parameter $\mu$, in a mixed computation with that in the full vertex
expansion stabilises the results in a particular matter gravity system.
Hence, this
gives us some trust in the qualitative results for higher $\R^n$ terms
in the background field approximation. In \cite{Falls:2013bv} the $f(\R)$-
potential has been computed polynomially up to $\R^{34}$, and
the relevance of these operators follows the perturbative
counting closely. Accordingly it is quite probable that the higher $\R^n$ will
turn out to be irrelevant in the full vertex expansion as well. 

Based on the above observations we have also constructed projection
operators that properly disentangle the contributions of different diffeomorphism-invariant tensor structures.
This allowed us to switch off the $\R^2$ coupling in order to analyse its importance for the system.
In this case, we are led to an unstable system, which highlights the importance of the $\R^2$ coupling for the
asymptotic safety scenario. In the present work we include the
$\R^2$ contributions via the momentum dependence of the gravitational
coupling $\gn{4}(p^2)$, leading to a very stable system in the UV. 

In the full system with $\R^2$ contributions we found an attractive UV fixed
point with three attractive directions and two repulsive directions.
The third attractive direction can be explained due to the overlap
with $\R^2$, and is in agreement with previous $\R^2$ studies in the
background field approximation
\cite{Lauscher:2002sq,Codello:2007bd,Machado:2007ea,Codello:2008vh,Falls:2013bv}.
We investigated the stability of this UV fixed point with respect to
changes of the identification of the higher couplings and compared it
to the stability of the previous truncation without the graviton
four-point function.  We characterised the stability via the area of
existence in the theory space of higher couplings, and remarkably this
area increased with the improved truncation.  We interpret this as a
sign that the systematic approximation scheme is approaching a converging
limit.

Furthermore, we investigated the IR behaviour and 
found trajectories that connect the UV fixed point with classical
general relativity. In particular, we found two different types of
such trajectories. In the first category all couplings, including
background and fluctuation couplings, scale classically according
to their mass dimension below the Planck scale. 
In consequence the Nielsen identities become trivial in this regime 
and we can solve them in the IR.
In the second category,
the graviton mass parameter and the coupling $\lambdan{3}$ scale
non-classically below the Planck scale, which is triggered by the
graviton mass parameter flowing towards the pole of the graviton
propagator \mbox{$\mu\to-1$}. In summary, the IR behaviour was found to be
very similar to \cite{Christiansen:2014raa}, and recently also \cite{Biemans:2016rvp}.

Lastly, we discussed signs of apparent convergence in the present
system by comparing the results to previous truncations.  As mentioned
before, we observed that the present system is more stable and less
sensitive to the closure of the flow equation, which is expected from
a converging system. We furthermore used two different approximations
of the stability matrix and argued that the critical exponents
belonging to the most attractive directions should not differ in a
well converged expansion. Indeed we found that the difference of the
critical exponents is decreasing with an improvement of the
truncation. We interpret this as a sign towards convergence.

In the present approximation we have taken the $\Lambda$,
$\R$ and $\R^2$ tensor structures into account.
Furthermore, we have shown that
the higher derivative tensor structures and the
$\RicciTensor_{\mu\nu}^2$ tensor structure are
suppressed. There are also very strong indications for the irrelevance
of the higher orders in $\R^n$. Altogether this suggests that the
natural extension of the present work towards apparent convergence
consists primarily of the inclusion of all tensor structures of the
vertices and propagators on the given level $n=4$ of the vertex
expansion. In particular, this concerns the inclusion of the
$\R^2$ tensor structure with the orthogonal projection devised in
the present work. Moreover, the different graviton modes should be
furnished with their separate dispersion or wave function
renormalisation. This is well in reach with the current technical
status of the programming code and its implementation.  Then, selected
tensor structures of higher vertices could be used for further tests
of apparent convergence.  We hope to report on this in the near
future.

\vspace{.5cm}
\noindent \textbf{Acknowledgements} We thank N.~Christiansen,
A.~Eichhorn, K.~Falls, S.~Lippoldt and A.~Rodigast for discussions. MR
acknowledges funding from IMPRS-PTFS. This work is supported by EMMI
and by ERC-AdG-290623.

\appendix

\section{Approximations of the stability matrix}\label{app:stability-matrix}
The stability matrix $\stabilityMatrix$ is defined as the Jacobi
matrix of the flow equations for all couplings $\couplingg{i}$.  Mathematically, it is
given by
\begin{align}
 \stabilityMatrix_{i j} \definition \partial_{\couplingg{j}} \couplinggd{i}\,.
\end{align}
In this work, the critical exponents of a fixed point are defined as the eigenvalues
of the stability matrix evaluated at this fixed point.  In our setup,
the stability matrix is infinite dimensional since it is spanned by
all couplings $\lambdan{n}$ and $\gn{n}(p^2)$.  Note that one momentum
dependent coupling alone would already be enough to render the stability
matrix infinite dimensional.

In the present work, only couplings up to order six appear in the
flow.  We furthermore do not resolve the full momentum dependence of
the couplings.  Thus, we have already rendered the stability matrix
finite.  Nevertheless, the full stability matrix is not known since
the flows of the fifth- and sixth order couplings are unknown and
would depend on further higher couplings anyway.  In consequence, we
have to make an approximation of the stability matrix in order to
obtain the critical exponents.  Note that we also have to make an
approximation of the flow itself in order to close it.  Naturally,
these approximations are related.

In the following we present two different approximations of the
stability matrix.  We further argue that these approximations should
give approximately the same values for the most relevant critical
exponents if the expansion scheme is already well converged.

The approximation of the flow is related to its closure and describes
how the higher couplings are identified with the lower ones.  In the
following, we call this process identification scheme and denote it by
$|_{\text{id.}}$.  The two different approximations of
the stability matrix are distinguished by the sequence of taking the
derivatives and applying this identification scheme.  In the first
approximation, the identification is performed \emph{before} taking
the derivatives:
\begin{align}
  {\bar \stabilityMatrix}_{i j} \definition \partial_{\couplingg{j}} (\couplinggd{i}|_{\text{id.}}) \,.
\end{align}
The critical exponents that correspond to this approximation represent
the critical exponents that belong to the computed phase diagram of
the theory.

In the second approximation, the identification is performed \emph{after} taking the derivatives, to wit
\begin{align}
 \stabilityMatrixAlt_{i j} \definition (\partial_{\couplingg{j}} \couplinggd{i}) |_{\text{id.}}\,.
\end{align}
This approximation is more closely related to the full stability matrix in the sense that it respects the fact that the higher couplings in the full system do not coincide with the lower ones.
Note that these two different approximations only differ if we choose a non-trivial identification scheme, i.e.\ if the higher couplings are functions of the lower ones.

If the expansion scheme is well converged, then the contributions of the higher couplings to the flow of the lower couplings are small, 
e.g.\ $(\partial_{\lambda_{n_{\text{max}}+2}} \lambda_{n_{\text{max}}})|_{\text{FP}}\approx 0$.
In this case, the most relevant eigenvalues of the stability matrices $\bar\stabilityMatrix$ and $\stabilityMatrixAlt$ coincide approximately.
The stabilisation of the most relevant eigenvalues was also observed in an expansion in $\R^n$ in~\cite{Falls:2013bv}.
In consequence, a huge deviation in the most relevant eigenvalues of both approximations would clearly indicate a lack of convergence.
For this reason, we use the comparison of the different approximations as a first check of convergence.

\begin{table*}
  \centering
  \caption[UV fixed points for different identification schemes]{
    Properties of the non-trivial UV fixed point for different identification schemes, 
    i.e.\ different closures of the flow equations, as discussed in App.~\ref{app:identification-scheme}.
    The flow equations are computed with momentum dependent anomalous dimensions $\eta_{\phi_i}$ and bilocally projected Newton's couplings $\gn{n}(k^2)$.
    The critical exponents $\bar \criticalExponent_i$ and $\tilde \criticalExponent_i$ 
    stem from two different approximation of the stability matrix as discussed in App.~\ref{app:stability-matrix}.
    An attractive UV fixed point is found in most identification schemes with mildly varying fixed point values.
    In the first approximation of the stability matrix we always find three attractive directions,
    while in the second approximation of the stability matrix we find one or three attractive directions,
    since the real part of one complex pair of eigenvalues is quite close to zero.
    These results suggest that the present system is rather stable under change of the closure of the flow equations.
    In the case of the single identification without a physical UV fixed point we found that it had in fact just vanished in the complex plane.
  }
  \label{tab:UV_fixed_points_identification_overview}
  \vspace{5pt}
  \setlength{\tabcolsep}{2.5pt}
  \begin{tabular}{ l @{\hskip 14pt} c  c  c  c  c  @{\hskip 14pt} c  c  c  c  c  @{\hskip 14pt} c  c  c  c  c }
    \toprule
    Identification scheme &  $\mu^*$ & $\lambdan{3}^*$ & $\lambdan{4}^*$ & $\gn{3}^*$ & $\gn{4}^*$ & \multicolumn{5}{c}{$\bar \criticalExponent_i$}  &  \multicolumn{5}{c}{$\tilde \criticalExponent_i$} \\
    \midrule
    $\gn{n>4}\to \gn{3}$, $\lambdan{n>4}\to \lambdan{3}$ & $-0.48$ & $0.092$ & $0.0077$ & $0.62$ & $0.53$ & $-5.0$ & \multicolumn{2}{c}{$-1.3 \pm 3.4\imaginaryi$} & $3.7$ & $10$ & $-4.2$ & \multicolumn{2}{c}{$0.62 \pm 1.8\imaginaryi$} & $4.7$ & $9.3$ \\
    $\gn{n>4}\to \gn{4}$, $\lambdan{n>4}\to \lambdan{3}$ & $-0.45$ & $0.12$ & $0.028$ & $0.83$ & $0.57$ & $-4.7$ & \multicolumn{2}{c}{$-2.0 \pm 3.1 \imaginaryi$} & $2.9$ & $8.0$ & $-5.0$ & \multicolumn{2}{c}{$-0.37 \pm 2.4 \imaginaryi$} &  $5.6$ & $7.9$ \\
    $\gn{n>4}\to \gn{4}$, $\lambdan{n>4}\to \lambdan{4}$ & \multicolumn{5}{l}{physical UV fixed point not found} \\
    $\gn{n>4}\to \gn{4}$, $\lambdan{6}\to \lambdan{4}$, $\lambdan{5}\to \lambdan{3}$ & $-0.49$ & $0.086$ & $0.027$ & $0.64$ & $0.56$ & $-8.7$ & \multicolumn{2}{c}{$-1.4 \pm 3.7 \imaginaryi$} & $4.3$ & $11$ & $-5.0$ & \multicolumn{2}{c}{$0.46 \pm 2.0 \imaginaryi$} &  $5.5$ & $11$ \\
    \bottomrule
  \end{tabular}
\end{table*}

\section{Background couplings} \label{app:background_couplings}
In this section we present the flow equations for the background couplings
$\backgroundg$ and $\backgroundLambda$.  They are in particular
interesting in the limit $k \to 0$ where they become observables.  In
this limit, the regulator term vanishes by construction and diffeomorphism invariance is restored, which implies that these
couplings can be interpreted as physical observables only for
vanishing $k$.

For notational convenience we reintroduce the coupling $\lambda_2 = -
\mu/2$.  Following~\cite{Gies:2015tca}, we compute the flow of the
background couplings with a curvature expansion on an Einstein space.
We use a York-decomposition~\cite{York:1973ia,Stelle:1976gc} and field
redefinitions~\cite{Dou:1997fg,Lauscher:2001ya} to cancel the
non-trivial Jacobians.  The resulting flow equations are given by
\begin{align}\label{eq:bg_flows_2}
  \partial_t {\bar g} &= 2 \bar g  - \bar g^2 f_{R^1}(\lambda_2;\eta_\phi)\,, \notag\\[2ex]
  \partial_t{\bar\lambda} &= -4 \bar \lambda + \bar \lambda \frac{\partial_t{\bar g}}{\bar g}+ \bar g f_{R^0}(\lambda_2;\eta_\phi)\,,
\end{align}
where the functions $f_{\R^0}$ and $f_{\R^1}$ read
\begin{widetext}
\begin{align}
  f_{\R^0}(\lambda_2;\eta_\phi) &= \frac{1}{24 \pi} \left(\frac{(10-8 \lambda_2)(6-\eta_h(k^2))}{1 - 2  \lambda_2}-8 (6-\eta_c(k^2))\right)\,,\notag\\[2ex]
  f_{\R^1}(\lambda_2;\eta_\phi) &= \frac{1}{24 \pi} \left(\frac{93+204
      \lambda_2-300 \lambda_2^2-\eta_h(k^2) \left(17+36 \lambda_2-60
        \lambda_2^2\right)}{3 (1 -2 \lambda_2)^2} + 10
    (5-\eta_c(k^2))\right) \,. \label{eq:fRi}
\end{align}
\end{widetext}
In consequence the fixed point equations for the background couplings
are given by
\begin{align}\label{eq:bg_flows_fp}
  \backgroundg^* &= \frac{2}{f_{\R^1}(\lambdan{2}^*,\eta_\phi^*)}\,, \notag\\[2ex]
  \backgroundLambda^* &= \frac{f_{\R^0}(\lambdan{2}^*,\eta_\phi^*)}{2
    f_{\R^1}(\lambdan{2}^*,\eta_\phi^*)}\,.
\end{align}
Note that the background couplings are non-dynamical, i.e.\ they do not
influence any other coupling.  Furthermore, the background couplings
only depend on the couplings of the two-point function.  Hence only
the graviton mass parameter $\mu$ (or equivalently $\lambda_2$) and
the anomalous dimensions $\eta_h$ and $\eta_c$ directly affect them.

\begin{figure*}[tbp]
  \centering
  \includegraphics[width = \textwidth]{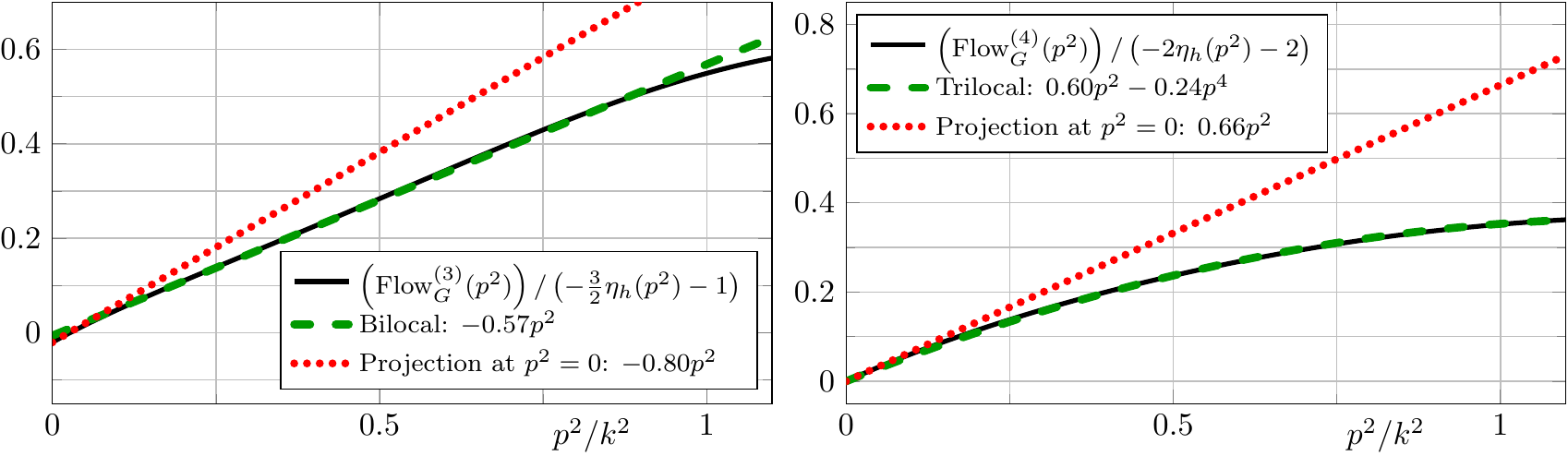}
  \caption[Momentum dependence with local momentum projection]
  {Fit of a local momentum projection at $p^2=0$ to the momentum
    dependence of the flow of the graviton three-point function (left)
    and the graviton four-point function (right) divided by
    $(-\frac{n}{2} \anomalousDimension{\graviton}(p^2) - n + 2)$ as
    defined in \eqref{eq:flow_polynomial}.  The flows are
    evaluated at $\left( \mu, \lambdan{3}, \lambdan{4}, \gn{3}, \gn{4}
    \right) = \left( -0.4, 0.1, -0.1, 0.7, 0.5 \right)$ and
    $\lambdan{6}=\lambdan{5}=\lambdan{3}$ as well as
    $\gn{6}=\gn{5}=\gn{4}$, i.e.\ the same values as in
    \autoref{fig:mom_dep_flow} where a non-local momentum projection was
    used.  In comparison to the non-local momentum projection, the
    local momentum projection does \textit{not} capture the correct
    momentum dependence in the whole momentum range $0\leq p^2 \leq
    k^2$ since it is sensitive to local momentum fluctuations.
    Furthermore, it is technically very challenging to project on the
    $p^4$-term due to IR singularities.  Note again that the constant
    parts of the flows are irrelevant for the beta functions since
    they are extracted from a different tensor projection.  }
  \label{fig:mom_dep_local_momentum_projection}
\end{figure*}

\section{Dependence on the identification scheme}
\label{app:identification-scheme}
The flow of each $n$-point function depends on the couplings of the $(n+1)$-point function and the $(n+2)$-point function, see \autoref{fig:flow_n_point}.
For the highest couplings, we consequently do not have a flow equation at hand.
In our setup, these are the couplings of the five- and six-point function, i.e.\ $\lambdan{5}$, $\lambdan{6}$, $\gn{5}$, and $\gn{6}$.
In order to close the flow of our system, we need to make an ansatz for these higher order couplings.
A natural choice is one that is close to diffeomorphism invariance, i.e.\ to identify these couplings with a lower order coupling.

In our setup, there are two lower order couplings that correspond to a (partly) diffeomorphism invariant identification scheme, e.g.\ $\lambdan{5}$ can be identified with $\lambdan{3}$ or $\lambdan{4}$.
In a well converged expansion scheme, the details of the identification should not matter and lead to similar results.
In this section, we compare the results for different identification schemes in order to evaluate the stability of our expansion scheme.

The properties of the non-trivial UV fixed point for different identifications schemes are displayed in \autoref{tab:UV_fixed_points_identification_overview}.
In all identification schemes except for the identification $\gn{n>4}\to \gn{4}$ and $\lambdan{n>4}\to \lambdan{4}$ we find an attractive UV fixed point.
In this case we can see that the fixed point has just vanished in the complex plane.
For all other identifications we observe that the fixed point values and the critical exponents vary only mildly.
Especially the number of attractive directions is consistently three with the first approximation to the stability matrix, c.f.\ App.~\ref{app:stability-matrix}.
With the second approximation the number of attractive directions varies from one to three since the real part of one complex conjugated pair is close to zero.

In conclusion, this analysis suggests that our system is rather stable with respect to different identification schemes.
Only one particular identification scheme has led to the disappearance of the attractive UV fixed point.
This constitutes further support for our results in \autoref{sec:stability}, where we found that our full system is very stable with respect to the identification of $\gn{5}$ and $\gn{6}$.

\section{Possible issues of a local momentum projection}
\label{app:local-momentum-projection}
In this section we want to point out some possible issues of a local momentum projection.
A local momentum projection is for example a derivative expansion about a certain momentum, usually $p=0$.

The full solution of a flow equation includes a full resolution of the momentum dependence of all vertex flows.
For higher $n$-point functions this task is computationally extremely challenging due to the high number of momentum variables.
We have already argued in \autoref{sec:flows-proj} that this task can be tremendously simplified with a symmetric momentum configuration.
We have further shown in \autoref{sec:mom-dep-nptfct} that the quantity $\flow^{(n)}_G/(-\frac{n}{2} \anomalousDimension{\graviton}(p^2) - n + 2)$  is polynomial in $p^2$, at least for $n=3,4$.
Thus it is possible to consistently project on each coefficient of this polynomial in the whole momentum range $0\leq p^2 \leq k^2$ by employing a non-local momentum projection.
In contrast, a local momentum projection scheme does not capture the correct momentum dependence over the whole momentum range $0\leq p^2 \leq k^2$ in general
since it is sensitive to local momentum fluctuations.
Furthermore, it is very challenging to project on the $p^4$ coefficient or even higher momentum order coefficients due to IR singularities. All these statements are explicitly exemplified in \autoref{fig:mom_dep_local_momentum_projection}.

On the other hand, the local momentum projection at $p=0$ has the advantage that it allows for analytic flow equations, as discussed in App.~\ref{app:analytic-eq}.
Analytic flow equations are more easily evaluated in the whole theory space,
but, as the discussion above suggests, one should be be mindful of the fact that they easily introduce a large error.

We use the analytic flow equations in \autoref{sec:IR-behaviour} precisely for the reason 
that they can easily be evaluated in the whole theory space.
Thus we show now that the fixed point properties in this analytic system are qualitatively similar to the full system, 
despite the error that is introduced by the analytic equations.

The properties of the UV fixed point for different approximations are displayed in \autoref{tab:UV_fixed_points_truncations_4pt}.
Truncation 1 corresponds to our full system, 
i.e.\ with momentum dependent anomalous dimensions and bilocally evaluated gravitational couplings.
Truncation 4 corresponds to the system used in \autoref{sec:IR-behaviour}, 
i.e.\ without anomalous dimensions and with gravitational couplings from a derivative expansion.
Truncation 2 and 3 are in between those truncations, 
i.e.\ with anomalous dimension but gravitational couplings from a derivative expansion 
and without anomalous dimensions but with bilocally evaluated gravitational couplings, respectively.

We observe that the UV fixed point exists in all truncations, 
and that the properties of this fixed point vary only mildly.
The fixed point values are all located within a small region, 
with the exception of our simplest truncation.
There, the couplings $\lambdan{3}$ and $\lambdan{4}$ have a different sign compared to the other truncations.
Considering the critical exponents we always find three attractive directions with the first approximation of the stability matrix
and in three out of four cases three attractive directions with the second approximation of the stability matrix as well.
These results suggest that it is an acceptable approximation to use the analytic flow equations 
if one is only interested in the qualitative behaviour of the system.

\begin{table*}
  \centering
  \caption[UV fixed points of the graviton four-point function]{
    Properties of the non-trivial UV fixed points for different approximations within the full system.
    In the truncations 3 and 4, we set $\anomalousDimension{\phi_i} = 0$, while in the truncations 1 and 2 we use momentum dependent anomalous dimensions.
    In truncations 2 and 4, the couplings $\gn{3,4}$ are computed via a derivative expansion at $p=0$,
    while in the truncations 1 and 3 the couplings $\gn{3,4}(k^2)$ are evaluated with a bilocal projection between $p=0$ and $p=k$.
    The quality of the truncation decreases from 1 to 4.
    The fixed point values are obtained with the identification scheme $\lambdan{n>4} = \lambdan{3}$ and $\gn{n>4} = \gn{4}$.
    The critical exponents $\bar \criticalExponent_i$ and $\tilde \criticalExponent_i$ stem from two different approximation of the stability matrix as discussed in App.~\ref{app:stability-matrix}.
    The fixed point properties from different approximations are qualitatively very similar.
    In particular, all fixed points exhibit three relevant directions when the first approximation of the stability matrix is used.
    Using the second approximation of the stability matrix also results in three relevant directions in three out of four cases.
    }
  \label{tab:UV_fixed_points_truncations_4pt}
  \vspace{5pt}
  \setlength{\tabcolsep}{5pt}
  \begin{tabular}{ l @{\hskip 15pt} c  c  c  c  c  @{\hskip 25pt} c  c  c  c  c  @{\hskip 25pt} c  c  c  c  c }
    \toprule
      Trunc. &  $\mu^*$ & $\lambdan{3}^*$ & $\lambdan{4}^*$ & $\gn{3}^*$ & $\gn{4}^*$ & \multicolumn{5}{c}{$\bar \criticalExponent_i$}  &  \multicolumn{5}{c}{$\tilde \criticalExponent_i$} \\
    \midrule
      1 & $-0.45$ & $0.12$ & $0.028$ & $0.83$ & $0.57$ & $-4.7$ & \multicolumn{2}{c}{$-2.0 \pm 3.1 \imaginaryi$} & $2.9$ & $8.0$ & $-5.0$ & \multicolumn{2}{c}{$-0.37 \pm 2.4 \imaginaryi$} & $5.6$ & $7.9$ \\
      2 & $-0.41$ & $0.076$ & $0.0055$ & $0.71$ & $0.53$ & $-4.0$ & \multicolumn{2}{c}{$-1.5 \pm 3.6 \imaginaryi$} & $3.1$ & $6.0$ & $-3.9$ & \multicolumn{2}{c}{$-0.38 \pm 4.4 \imaginaryi$} & $2.3$ & $6.6$ \\
      3 & $-0.37$ & $0.049$ & $0.0055$ & $1.1$ & $0.83$ & $-7.3$ & \multicolumn{2}{c}{$-1.7 \pm 2.1 \imaginaryi$} & $3.0$ & $6.8$ & $-7.2$ & \multicolumn{2}{c}{$0.32 \pm 2.7 \imaginaryi$} &  $4.7$ & $6.8$ \\
      4 & $-0.23$ & $-0.060$ & $-0.11$ & $0.64$ & $0.55$  & $-3.0$ & \multicolumn{2}{c}{$-1.9 \pm 1.6 \imaginaryi$} & $1.7$ & $3.4$ & $-2.2$ & \multicolumn{2}{c}{$-0.50 \pm 1.7 \imaginaryi$} &  \multicolumn{2}{c}{$1.5 \pm 0.88 \imaginaryi$} \\
    \bottomrule
  \end{tabular}
\end{table*}

\section{Derivation of flow equations }\label{app:derivation-eq}
We obtain the flow equations for the individual coupling constants by projecting onto the flow of the graviton \nptfct{n}s, as explained in \autoref{sec:flows-couplings}.

The equations for the graviton mass parameter $\mass$ and for the graviton anomalous dimension $\anomalousDimension{\graviton}$ 
are extracted from the transverse-traceless part of the flow of the graviton two-point function.
For $p^2 = 0$, we obtain
\begin{align} \label{eq:mu_dot}
  \scaleDerivative{\mass} =  \left( \anomalousDimension{\graviton}(0) - 2 \right) \mass  + \0{32 \pi}{5} \flow^{(\graviton \graviton)}_\ttSymb (0) \,,
\end{align}
for the flow of the graviton mass parameter.

We obtain an equation for the graviton anomalous dimension $\anomalousDimension{\graviton}(p^2)$ by evaluating the flow of the graviton two-point function bilocally at $p^2$ and $-\mass k^2$, to wit
\begin{align} \label{eq:anomalous_dimension_grav_bilocal}
  \anomalousDimension{\graviton}(p^2) =  \0{32 \pi}{5(p^2 + \mass k^2)	} &\left(\flow^{(\graviton\graviton)}_{\ttSymb} \left(-\mass k^2 \right)\right. \notag \\
  &\hspace{1cm}-\left. \flow^{(\graviton\graviton)}_{\ttSymb} \left( p^2 \right) \right) \,.
\end{align}
The ghost anomalous dimension $\anomalousDimension{\ghost}(p^2)$ can be directly obtained from the transverse flow of the ghost two-point function, to wit
\begin{align}
  \anomalousDimension{\ghost}(p^2) &= - \frac{ \flow^{(\antighost\ghost)}_{\tSymb} (p^2)}{3 p^2} \label{eq:anomalous_dimension_ghost}\,.
\end{align}

In case of the higher order couplings, we employ the projection operators described in \autoref{sec:flows-proj}.
For the couplings $\lambdan{n}$, this leads to
\begin{align}
    \lambdand{n} &= \left(\frac{n}{2}\anomalousDimension{\graviton}(0) +(n-4) -\frac{n-2}{2}\frac{\gnd{n}}{\gn{n}} \right) \lambdan{n} \notag\\[2ex]
    &\phantom{=\,} + \frac{\gn{n}^{1-\frac{n}{2}}}{\constantLambdan{n}} \flow^{(n)}_{\Lambdan{n}}(0) \,, \label{eq:lamn_dot}
\end{align}
where the projection dependent constant $\constantLambdan{n}$ is implicitly defined via $\projLambdan{n} \contract \projTT^n \contract \tensorStructure^{(n)}(0;1) = \constantLambdan{n}$.
Here, $\contract$ denotes the pairwise contraction of indices.

As discussed in \autoref{sec:flows-couplings}, the gravitational couplings $\gn{n}(p^2)$ are momentum dependent.
In order to simplify the computation we make an approximation of the full momentum dependence.
This approximation exploits the fact that the flows are peaked at $p^2 = k^2$ and 
consequently we set the feed back on the right-hand side of the flow equation to $\gn{n}(p^2) \approx \gn{n}(k^2)$.
This closes the flow equation for $\gn{n}(k^2)$ and thus we only solve this equation.
The easiest way to obtain the flow equation for $\gn{n}(k^2)$ is a bilocal projection at $p=0$ and $p=k$.
The resulting equation for $g_3(k^2)$ is precisely the same as in \cite{Christiansen:2015rva}.
For $g_4(k^2)$ we obtain
\begin{widetext}
\begin{align} \label{eq:g4-k2-flow}
 \scaleDerivative{\gn{4}}(k^2) &= 2 \gn{4}(k^2) + 2 \anomalousDimension{\graviton}(k^2)\gn{4}(k^2) - C_4 \gn{4}(k^2)\lambdan{4} (\anomalousDimension{\graviton}(k^2) - \anomalousDimension{\graviton}(0))  + {\constantGnp{4}}^{-1}(\flow_G^{(4)}(k^2)- \flow_G^{(4)}(0))\,.
\end{align}
\end{widetext}
The derivation of this equation is based on the assumption that $\lambdan{4}$ is small.

In \autoref{sec:R2-coupling} we have laid out a strategy to disentangle contributions from different tensor structures, in particular those of $R$ and $R^2$.
The flow equations for the $g_n$ are obtained by a projection onto the $p^2$ part of $\flow^{(n)}_G$ divided by $\left(-\frac{n}{2}\anomalousDimension{\graviton}\left(p^2\right) - n + 2 \right)$, see \autoref{sec:mom-dep-nptfct} and \autoref{sec:R2-coupling}.
The graviton three-point function is at most quadratic in the external momentum, and consequently it is again enough to use a bilocal projection at $p^2 = 0$ and $p^2 = k^2$.
Consequently, the flow equation for $g_3$ is quantitatively equivalent to the previous one if $\lambdan{3}$ is small.
The graviton four-point function, on the other hand, has $p^4$ as its highest momentum power, and thus we use a trilocal momentum projection at $p^2=0$, $p^2=k^2/2$, and $p^2=k^2$.
The flow equations of $\gn{3}$ and $\gn{4}$ are then given by
\begin{widetext}
\begin{align} \label{eq:g3_flow}
    \left(1 + \anomalousDimension{3}\right)  \scaleDerivative{\gn{3}} &= 2 \gn{3} - 2 \gn{3} C_3 \left(\scaleDerivative{\lambdan{3}} +2 \lambdan{3}\right)  \left( \cfrac{1}{\frac{3}{2}\anomalousDimension{\graviton}(k^2) + 1} - \cfrac{1}{\frac{3}{2}\anomalousDimension{\graviton}(0) + 1} \right) + \cfrac{2}{\constantGnp{3}\sqrt{\gn{3}}} \left( \cfrac{\flow_G^{(3)}(k^2)}{\frac{3}{2}\anomalousDimension{\graviton}(k^2) + 1} - \cfrac{\flow_G^{(3)}(0)}{\frac{3}{2}\anomalousDimension{\graviton}(0) + 1} \right)\,,  \\[2ex]
    \left(1 + \anomalousDimension{4}\right) \scaleDerivative{\gn{4}} &= 2 \gn{4} - \gn{4} C_4 \left(\scaleDerivative{\lambdan{4}} +2 \lambdan{4}\right)  \left( - \cfrac{1}{\anomalousDimension{\graviton}(k^2) + 1}+ \cfrac{4}{\anomalousDimension{\graviton}(k^2/2) + 1} - \cfrac{3}{\anomalousDimension{\graviton}(0) + 1}   \right) \nonumber \\[2ex]
    &\phantom{=\,} + \cfrac{1}{\constantGnp{4}}\left( - \cfrac{\flow_G^{(4)}(k^2)}{\anomalousDimension{\graviton}(k^2) + 1} + 4 \cfrac{\flow_G^{(4)}(k^2/2)}{\anomalousDimension{\graviton}(k^2/2) + 1} - 3 \cfrac{\flow_G^{(4)}(0)}{\anomalousDimension{\graviton}(0) + 1}  \right) \,,
    \label{eq:g4_flow}
\end{align}
\end{widetext}
where
\begin{align} \nonumber
    \anomalousDimension{3} &= \cfrac{C_3 \lambdan{3} - \frac{3}{2}\anomalousDimension{\graviton}(k^2)}{\frac{3}{2}\anomalousDimension{\graviton}(k^2) + 1} - \cfrac{C_3 \lambdan{3}}{\frac{3}{2}\anomalousDimension{\graviton}(0) + 1} \,, \\[2ex]
    \anomalousDimension{4} &=  \cfrac{-3 C_4 \lambdan{4}}{\anomalousDimension{\graviton}(0) + 1} + 4\cfrac{C_4 \lambdan{4} -\frac{1}{2}\anomalousDimension{\graviton}(k^2/2)}{\anomalousDimension{\graviton}(k^2/2) + 1}  - \cfrac{C_4 \lambdan{4} -\anomalousDimension{\graviton}(k^2)}{\anomalousDimension{\graviton}(k^2) + 1}  \,.
    \nonumber
\end{align}
The constants $C$ are implicitly defined via $\projGn{n} \contract \projTT^n \contract \tensorStructure^{(n)}(p^2;\Lambdan{n}) = \constantGnLambda{n} \Lambda_n +\constantGnp{n} p^2$,
and we use the abbreviation $C_n \definition \constantGnLambda{n}/\constantGnp{n}$.
Again, $\contract$ denotes the pairwise contraction of indices.
Note that the constants $\anomalousDimension{n}$ are chosen in such a way that $\anomalousDimension{n} = 0$ for vanishing anomalous dimensions.

Analogously, we can obtain a flow equation for the $R^2$ coupling of the graviton four-point function $\omega_4$ by using a trilocal momentum projection, as explained in \autoref{sec:R2-coupling}.
We evaluate the flows at the same momenta as for the trilocal flow equation of $\gn{4}$.
The equation for $\omega_4$ then reads
\begin{widetext}
\begin{align} \label{eq:o4_flow}
    \left(1 + \anomalousDimension{\omega} \right) \scaleDerivative{\RsquaredCouplingn{4}} = 2 \RsquaredCouplingn{4}  & - \frac{\scaleDerivative{\gn{4}}}{\gn{4}} \left( \cfrac{\constantGnLambda{4} \lambdan{4} + \constantGnp{4} + \constantGnRsquared{4} \RsquaredCouplingn{4}}{\anomalousDimension{\graviton}(k^2) + 1}  - 2\cfrac{\constantGnLambda{4} \lambdan{4} + \frac{1}{2} \constantGnp{4} + \frac{1}{4} \constantGnRsquared{4} \RsquaredCouplingn{4}}{\anomalousDimension{\graviton}(k^2/2) + 1} + \cfrac{\constantGnLambda{4} \lambdan{4}}{\anomalousDimension{\graviton}(0) + 1} \right) \nonumber \\[2ex]
    &- \cfrac{\constantGnLambda{4}}{\constantGnRsquared{4}} \left(\scaleDerivative{\lambdan{4}} + 2\lambdan{4} \right) \left( \cfrac{1}{\anomalousDimension{\graviton}(k^2) + 1}- \cfrac{2}{\anomalousDimension{\graviton}(k^2/2) + 1} +  \cfrac{1}{\anomalousDimension{\graviton}(0) + 1}  \right) \nonumber \\[2ex]
    &+ \frac{1}{\constantGnRsquared{4} \gn{4}} \left(\cfrac{\flow_G^{(4)}(k^2)}{\anomalousDimension{\graviton}(k^2) + 1} - 2 \cfrac{\flow_G^{(4)}(k^2/2)}{\anomalousDimension{\graviton}(k^2/2) + 1} + \cfrac{\flow_G^{(4)}(0)}{\anomalousDimension{\graviton}(0) + 1} \right)  \,,
\end{align}
\end{widetext}
where
\begin{align}
  \anomalousDimension{\omega} &=  \cfrac{\anomalousDimension{\graviton}(k^2/2)}{\anomalousDimension{\graviton}(k^2/2) + 1} - \cfrac{2\anomalousDimension{\graviton}(k^2)}{\anomalousDimension{\graviton}(k^2) + 1} \,. \nonumber
\end{align}
The constants $C$ are again defined via the contraction $\projGn{n} \contract \projTT^n \contract \tensorStructure^{(n)}(p^2;\Lambdan{n}) = \constantGnLambda{n} \Lambda_n + \constantGnp{n} p^2+\constantGnRsquared{n} \Omega_4 p^4$.
Again, $\anomalousDimension{\omega}$ is defined such that $\anomalousDimension{\omega}=0$ for vanishing anomalous dimensions.

In the previous paragraphs we introduced abbreviations for constants that arise from the projection scheme.
The explicit values of these constants are:
\begin{align}
  \constantLambdan{3} &= \frac{5}{192 \pi^2}\,, & \constantLambdan{4} &= \frac{371881}{6718464\pi^2}\,, \nonumber \\[1ex]
  \constantGnLambda{3} &=  -\frac{9}{4096 \pi^2}\,, & \constantGnLambda{4} &= \frac{222485}{60466176\pi^2}\,, \nonumber \\[1ex]
  \constantGnp{3} &= \frac{171}{32768 \pi^2}\,, & \constantGnp{4} &= \frac{6815761}{544195584\pi^2}\,, \nonumber \\[1ex]
  C_3 &= \frac{\constantGnLambda{3}}{\constantGnp{3}} =  - \frac{8}{19}\,, & C_4 &= \frac{\constantGnLambda{4}}{\constantGnp{4}} = \frac{2002365}{6815761}\,, \nonumber \\[1ex]
  & & \constantGnRsquared{4} &=  -\frac{96203921}{1632586752\pi^2}\,.
\end{align}

If we want to obtain analytic flow equations for the gravitational couplings $\gn{n}$, 
which are significantly less accurate, as discussed in App.~\ref{app:local-momentum-projection}, 
then we have to apply a partial derivative with respect to $p^2$ and evaluate the result at $p^2=0$. 
The resulting equations are given by
\begin{align}
  \scaleDerivative{g}_n &= 2 g_n+ \frac{n g_n}{n-2} \left(\anomalousDimension{\graviton}(0) + C_n \lambda_n \anomalousDimension{\graviton}'(0)\right)   \nonumber \\[2ex]
  &\phantom{= \;} +\frac{2}{n-2}\frac{\gn{n}^{2-\frac{n}2}}{\constantGnp{n}} {\flow^{(n)}_{G}}'(0) \,, \label{eq:gn-analytic}
\end{align}
where $'$ denotes the dimensionless derivative with respect to $p^2$.
These equations remain completely analytic if we use a Litim-shaped regulator \cite{Litim:2000ci} and 
approximate the anomalous dimensions as constant, $\anomalousDimension{\phi}\left(q^2\right) \approx \text{const.}$.
We present the explicit analytic flow equations in the next Appendix.

\section{Analytic flow equations}\label{app:analytic-eq}
All analytic flow equations are derived at $p^2 = 0$ (see e.g.\ \eqref{eq:gn-analytic}) and with a Litim-shaped regulator \cite{Litim:2000ci}. 
The anomalous dimensions in the momentum integrals are approximated as constant, 
i.e.\ $\anomalousDimension{\phi_i} (q^2) \approx \anomalousDimension{\phi_i}$. 
It is usually a good approximation to use the anomalous dimensions evaluated at $k^2$ \cite{Meibohm:2015twa}. 
The analytic flow equations are then given by
\begin{widetext}
\begin{align}
  \scaleDerivative{\mu} =& \left( \anomalousDimension{\graviton}(0) - 2 \right) \mu
  \\& + \frac{1}{12 \pi}\frac{\gn{4}}{(1+\mu)^2} \left( 3 (\anomalousDimension{\graviton} - 8) - 8 \lambdan{4} (\anomalousDimension{\graviton}-6) \right) \nonumber \\
  &- \frac{1}{180 \pi}\frac{{\gn{3}}}{(1+\mu)^3} \left( 21 (\anomalousDimension{\graviton} - 10) - 120 \lambdan{3} (\anomalousDimension{\graviton} - 8) +320 \lambdan{3}^2 (\anomalousDimension{\graviton} - 6) \right) \nonumber \\
  &+ \frac{\gn{3}}{5 \pi} (\eta_{\ghost} - 10) \nonumber \\
  \scaleDerivative{\lambda}_3 =& \left(\frac{3}{2} \anomalousDimension{\graviton}(0) - 1 -\frac{1}{2} \frac{\scaleDerivative{g}_3}{\gn{3}}\right) \lambdan{3}
  \\& - \frac{1}{8 \pi}\frac{{\gn{3}}^{-\frac{1}{2}} {\gn{5}}^{\frac{3}{2}}}{(1+\mu)^2} \left((\anomalousDimension{\graviton} - 8) - 4 \lambdan{5} (\anomalousDimension{\graviton}-6) \right) \nonumber \\
  &- \frac{1}{6 \pi}\frac{{\gn{4}}}{(1+\mu)^3} \left( 3 \lambdan{4} (\anomalousDimension{\graviton} - 8) - 16 \lambdan{3} \lambdan{4} (\anomalousDimension{\graviton} - 6) \right) \nonumber \\
  &+ \frac{1}{240 \pi}\frac{ {\gn{3}}}{(1+\mu)^4} \left( 11 (\anomalousDimension{\graviton} - 12) - 72 \lambdan{3} (\anomalousDimension{\graviton} - 10) + 120 {\lambdan{3}}^2 (\anomalousDimension{\graviton} - 8) - 80 {\lambdan{3}}^3 (\anomalousDimension{\graviton} - 6) \right) \nonumber \\
  &- \frac{\gn{3}}{10 \pi} (\eta_{\ghost} - 12) \nonumber \\
  \scaleDerivative{\lambda}_4 =& \left(2 \anomalousDimension{\graviton}(0) - \frac{\scaleDerivative{g}_4}{\gn{4}}\right) \lambdan{4} + \frac{1}{13387716\pi} \Bigg(
  \\&\phantom{+\cdot\,} \frac{1}{2}\frac{{\gn{6}}^2 {\gn{4}}^{-1}}{(1+\mu)^2} \left(-4472787 (\anomalousDimension{\graviton} - 8) + 1639004 \lambdan{6} (\anomalousDimension{\graviton}-6) \right) \nonumber \\
  &+ \frac{1}{15}\frac{{\gn{4}}}{(1+\mu)^3} \left( 5066361 (\anomalousDimension{\graviton} - 10) - 22517160 \lambdan{4} (\anomalousDimension{\graviton} - 8) + 283174360 {\lambdan{4}}^2 (\anomalousDimension{\graviton} - 6) \right) \nonumber \\
  &+ \frac{2}{15} \frac{{\gn{3}}^{\frac{1}{2}} {\gn{5}}^{\frac{3}{2}} {\gn{4}}^{-1}}{(1+\mu)^3} \left( 3940503 (\anomalousDimension{\graviton} - 10) - 60 ( 187643 \lambdan{3} - 1303286 \lambdan{5})(\anomalousDimension{\graviton} - 8) + 417051520 \lambdan{3} \lambdan{5} (\anomalousDimension{\graviton} - 6) \right) \nonumber \\
  &+ \frac{2}{5} \frac{\gn{3}}{(1+\mu)^4} \left( - 1313501 (\anomalousDimension{\graviton} - 12) + 3377574 (2 \lambdan{3} + \lambdan{4}) (\anomalousDimension{\graviton} - 10) \nonumber \right.
  \\&\phantom{+\cdot\,} \left. - 15011440 (\lambdan{3} + 2 \lambdan{4}) \lambdan{3} (\anomalousDimension{\graviton} - 8) + 45442920 {\lambdan{3}}^2 \lambdan{4} (\anomalousDimension{\graviton} - 6) \right) \nonumber \\
  &+ \frac{1}{5}\frac{ {\gn{3}}^2 {\gn{4}}^{-1} }{(1+\mu)^5} \left( 2874147 (\anomalousDimension{\graviton} - 14) - 20879816 \lambdan{3} (\anomalousDimension{\graviton} - 12)  + 36027456 {\lambdan{3}}^2 (\anomalousDimension{\graviton} - 10) \nonumber \right.
  \\&\phantom{+\cdot\,} \left. + 88161840 {\lambdan{3}}^3 (\anomalousDimension{\graviton} - 8)  -248160672 {\lambdan{3}}^4 (\anomalousDimension{\graviton} - 6) \right) \nonumber \\
  &- \frac{10426288}{7} \frac{{\gn{3}}^2}{\gn{4}}(\eta_{\ghost} - 14) \Bigg) \nonumber \\
  \gnd{3} =& \left( 2 + 3 \anomalousDimension{\graviton}(0) - \frac{8}{19} \anomalousDimension{\graviton}'(0) \lambdan{3} \right)  \gn{3} + \frac{1}{19 \pi} \Big(
  \\& \phantom{+\cdot\,}\frac{{\gn{3}}^{\frac{1}{2}} {\gn{5}}^{\frac{3}{2}}}{(1+\mass)^2} \frac{47}{6}  (\anomalousDimension{\graviton} - 6) \nonumber
  \\& + \frac{\gn{3} \gn{4}}{18(1+\mass)^3} \left( - 45 (\anomalousDimension{\graviton} - 8) + 8 \left( 30 \lambdan{3}  - 59 \lambdan{4} \right) (\anomalousDimension{\graviton}-6)+ 360 \lambdan{3} \lambdan{4} (\anomalousDimension{\graviton}-4) \right) + \frac{\gn{3} \gn{4}}{(1+\mass)^4} 16 \left( 1 - 3 \lambdan{3} \right) \lambdan{4} \nonumber
  \\& - \frac{{\gn{3}}^2}{80(1+\mass)^4} \left( 147 (\anomalousDimension{\graviton}-10) - 1860 \lambdan{3} (\anomalousDimension{\graviton}-8) + 3380 \lambdan{3}^2 (\anomalousDimension{\graviton}-6) + 25920 \lambdan{3}^3 (\anomalousDimension{\graviton}-4) \right) \nonumber
  \\&\phantom{+\cdot\,} - \frac{2 {\gn{3}}^2}{15(1+\mass)^5} \left( 229 - 1780 \lambdan{3} + 3640 \lambdan{3}^2 - 2336 \lambdan{3}^3 \right) \nonumber
  \\& + \frac{{\gn{3}}^{2}}{10} \left( 53 (\anomalousDimension{\ghost}-10) + 480 \right) \Big) \label{eq:g3_derivative_analytic}  \nonumber 
\end{align}
\begin{align}
  \gnd{4} =& 2 \left(1 + \anomalousDimension{\graviton}(0) + \frac{2002365}{6815761} \anomalousDimension{\graviton}'(0) \lambdan{4} \right)  \gn{4} + \frac{2125764}{6815761 \pi} \Bigg(
  \\& \phantom{+\cdot\,}\frac{{\gn{6}}^{2}}{(1+\mass)^2} \frac{32830375}{25509168}  (\anomalousDimension{\graviton} - 6) \nonumber
  \\& -\frac{{\gn{4}}^2}{76527504(1+\mass)^3 } \left(11305705 (\anomalousDimension{\graviton} - 8) + 61298276 \lambdan{4} (\anomalousDimension{\graviton}-6) + 308793960 \lambdan{4}^2 (\anomalousDimension{\graviton}-4) \right) \nonumber
  \\&\phantom{+\cdot\,} -\frac{4 {\gn{4}}^2}{3188646(1+\mass)^4} \left(16061481 + 8 \left(5355213 \lambdan{4} -5610604\right) \lambdan{4} \right) \nonumber
  \\& -\frac{{\gn{3}}^{\frac{1}{2}} {\gn{5}}^{\frac{3}{2}}}{19131876(1+\mass)^3} \Big(-34242339 (\anomalousDimension{\graviton} - 8) + \left(86256922\lambdan{3} - 7511302\lambdan{5}\right) (\anomalousDimension{\graviton}-6) -4483422 \lambdan{3}\lambdan{5} (\anomalousDimension{\graviton}-4) \Big) \nonumber
  \\&\phantom{+\cdot\,} - \frac{{\gn{3}}^{\frac{1}{2}} {\gn{5}}^{\frac{3}{2}}}{(1+\mass)^4} \left(784609\left(17-32\lambdan{3}\right) - 8937232 (4-9 \lambdan{3}) \lambdan{5} \right) \nonumber
  \\& +\frac{{\gn{3}} {\gn{4}}}{90(1+\mass)^4} \Big(323831781 (\anomalousDimension{\graviton} - 10) -\left(80\cdot11179796\lambdan{3} + 203187860\lambdan{4}\right) (\anomalousDimension{\graviton}-8) \nonumber
  \\&\phantom{+\cdot\,}\phantom{+\cdot\,} -\left(10\cdot129631943\lambdan{3} - 80\cdot16941407\lambdan{4}\right) \lambdan{3} (\anomalousDimension{\graviton}-6) -10\cdot292902984 \lambdan{3}^2\lambdan{4}(\anomalousDimension{\graviton}-4)\Big) \nonumber 
  \\&\phantom{+\cdot\,} + \frac{{\gn{3}} {\gn{4}}}{15(1+\mass)^5} \Big(26769135 \left(17+8 \lambdan{3} \left(9 \lambdan{3} - 8\right)\right) -2 \left(353519805 + 4 \lambdan{3} \left(742510961 \lambdan{3} -514449355\right)\right) \lambdan{4}\Big) \nonumber 
  \\& +\frac{2{\gn{3}} {\gn{4}}}{9(1+\mass)^4} \Big(6783386859 (\anomalousDimension{\graviton} - 10) -\left(140\cdot86837935\lambdan{3} + 457106270\lambdan{4}\right) (\anomalousDimension{\graviton}-8) \nonumber
  \\&\phantom{+\cdot\,}\phantom{+\cdot\,} -\left(20\cdot4067950507\lambdan{3} + 140\cdot799593508\lambdan{4}\right) \lambdan{3} (\anomalousDimension{\graviton}-6) -20\cdot25136284404 \lambdan{3}^2\lambdan{4}(\anomalousDimension{\graviton}-4)  \Big) \nonumber 
  \\&\phantom{+\cdot\,} + \frac{{\gn{3}} {\gn{4}}}{15(1+\mass)^5} \Big(394709295 - 661068650 \lambdan{4} + 40 \left(91735671 \lambdan{4} -34781816\right)\lambdan{3} \nonumber
  \\&\phantom{+\cdot\,}\phantom{+\cdot\,}\phantom{+\cdot\,} - 8 \left(731880777 \lambdan{4} -220800215\right) \lambdan{3}^2\Big) \nonumber
  \\& +\frac{{\gn{3}}^2}{45(1+\mass)^5} \Big(-125220803(\anomalousDimension{\graviton}-12) +2\cdot284391180\lambdan{3}(\anomalousDimension{\graviton}-10) \nonumber
  \\&\phantom{+\cdot\,}\phantom{+\cdot\,} +2\cdot167456175\lambdan{3}^2(\anomalousDimension{\graviton}-8) -2\cdot236\cdot13640606\lambdan{3}^3(\anomalousDimension{\graviton}-6) + 2\cdot236\cdot36717495 \lambdan{3}^4(\anomalousDimension{\graviton}-4) \Big) \nonumber
  \\&\phantom{+\cdot\,} - \frac{{\gn{3}}^2}{3(1+\mass)^6} \Big(112533531 -855576992\lambdan{3} +3683259968\lambdan{3}^2 -7947008128\lambdan{3}^3 +6385327072\lambdan{3}^4\Big)\nonumber
  \\& +4 {\gn{3}}^{2} \left( \frac{23005837}{5} (\anomalousDimension{\ghost}-12) + 11171540 \right) \Bigg) \nonumber
    \label{eq:g4_flow_equation_derivative_analytic}
\end{align}
\end{widetext}

\bibliography{BibFile}

\end{document}